\documentclass[pdflatex,sn-mathphys-num]{sn-jnl}

\usepackage{graphicx}%
\usepackage{multirow}%
\usepackage{amsmath,amssymb,amsfonts}%
\usepackage{amsthm}%
\usepackage{mathrsfs}%
\usepackage[title]{appendix}%
\usepackage{xcolor}%
\usepackage{textcomp}%
\usepackage{manyfoot}%
\usepackage{booktabs}%
\usepackage{algorithm}%
\usepackage{algorithmicx}%
\usepackage{algpseudocode}%
\usepackage{listings}%
\usepackage{tabularx}
\usepackage{pifont}
\usepackage{rotating}
\usepackage{tikz}
\usepackage{transparent} 
\usepackage{wrapfig} 
\usepackage{enumitem}
\usepackage{silence}
\usepackage{caption}

\begin{document}

\title[Article Title]{Explainable AI-based Intrusion Detection System for Industry 5.0: An Overview of the Literature, associated Challenges, the existing Solutions, and Potential Research Directions}


\author*[1,4]{\fnm{Naseem} \sur{Khan}}\email{nakh12498@hbku.edu.qa}

\author[2]{\fnm{Kashif} \sur{Ahmad}}\email{kashif.ahmad@mtu.ie}
\equalcont{These authors contributed equally to this work.}

\author[4]{\fnm{Aref Al} \sur{Tamimi}}\email{altamimi@hbku.edu.qa}
\equalcont{These authors contributed equally to this work.}

\author[3]{\fnm{Mohammed M.} \sur{Alani}}\email{m@alani.me}
\equalcont{These authors contributed equally to this work.}

\author[1]{\fnm{Amine} \sur{Bermak}}\email{abermak@hbku.edu.qa}
\equalcont{These authors contributed equally to this work.}

\author[4]{\fnm{Issa} \sur{Khalil}}\email{ikhalil@hbku.edu.qa}
\equalcont{These authors contributed equally to this work.}

\affil*[1]{\orgdiv{Computer Science and Engineering}, \orgname{Hamad bin Khalifa University}, \state{Doha}, \country{Qatar}}

\affil[2]{\orgdiv{Computer Science}, \orgname{Munster Technological University}, \state{Cork}, \country{Ireland}}

\affil[3]{\orgdiv{Cybersecurity Research Lab}, \orgname{Toronto Metropolitan University}, \state{Toronto}, \country{Canada}}

\affil[4]{\orgdiv{Qatar Computing Research Institute}, \orgname{Hamad bin Khalifa University}, \state{Doha}, \country{Qatar}}


\abstract{Industry 5.0, which focuses on human and Artificial Intelligence (AI) collaboration for performing different tasks in manufacturing, involves a higher number of robots, Internet of Things (IoTs) devices and interconnections, Augmented/Virtual Reality (AR), and other smart devices. The huge involvement of these devices and interconnection in various critical areas, such as economy, health, education and defense systems, poses several types of potential security flaws. AI itself has been proven a very effective and powerful tool in different areas of cybersecurity, such as intrusion detection, malware detection, and phishing detection, among others. Just as in many application areas, cybersecurity professionals were reluctant to accept black-box ML solutions for cybersecurity applications. This reluctance pushed forward the adoption of eXplainable Artificial Intelligence (XAI) as a tool that helps explain how decisions are made in ML-based systems.  In this survey, we present a comprehensive study of different XAI-based intrusion detection systems for industry 5.0, and we also examine the impact of explainability and interpretability on Cybersecurity practices through the lens of Adversarial XIDS (Adv-XIDS) approaches. Furthermore, we analyze the possible opportunities and challenges in XAI cybersecurity systems for industry 5.0 that elicit future research toward XAI-based solutions to be adopted by high-stakes industry 5.0 applications. We believe this rigorous analysis will establish a foundational framework for subsequent research endeavors within the specified domain.}




\maketitle

\section{Introduction}\label{sec1}
The growing applications of Artificial Intelligence (AI) and Machine Learning (ML) have increased the need for a better understanding of AI-based solutions for smart industries, especially in Industry 5.0 applications. Similar to other sensitive application domains, such as business, healthcare, education, and defense systems, the enigmatic and obscure nature of AI raises concerns and the need for in-depth evaluation of the decisions made by these black box models in smart industries \cite{speith2022review}. In addition to issues of user rights, and intelligent technology acceptance, developers of these systems need to ensure the fair and unbiased nature of their solutions. The need to comprehend and interpret the causal understanding of inferences made by Deep ML models, directed the attention of the research community towards XAI  \cite{holzinger2022explainable}. In this regard, the first DARPA-funded Explainable AI (XAI) initiative started with the aims to develop interpretable machine learning models for reliable and human-trusted decision-making systems, crucial for the integration of Internet-of-Things (IoT) and intelligent systems in Industry 5.0 \citep{gunning2019darpa, alexandrov2017explainable, 9877919, yayla2022explainable}.

Cybersecurity is one of the critical aspects of smart industries involving a high number of interconnected devices. Similar to other applications, AI-based solutions have been proven very effective in the cybersecurity of smart industries. However, the opacity of the diploid complex AI-based models in various Cybersecurity solutions, such as Intrusion Detection Systems (IDS), Malware detection and classification systems, finding Zero-Day vulnerabilities, and Digital Forensics, exacerbates the trust lack of transparency  \citep{ 9877919, scalas2023improving}. To keep themselves one step ahead of attackers, it is also essential for security analysts to know the internal automatic decision mechanism of the deployed intelligent model and to precisely reason the input data about the model's outputs. The application of XAI in cybersecurity could also be a double-edged sword, that is, besides improving the security practices, it could also make the intelligent explainable model vulnerable to adversary attacks \cite{Marino2018AnAA, baniecki2023adversarial, sharma2022explainable}. Thus, the integration of human understanding and AI-based security systems needs to be keenly analyzed to provide a clear perception for future research.

\begin{figure*}[h!]
\centering
\includegraphics[width=\linewidth]{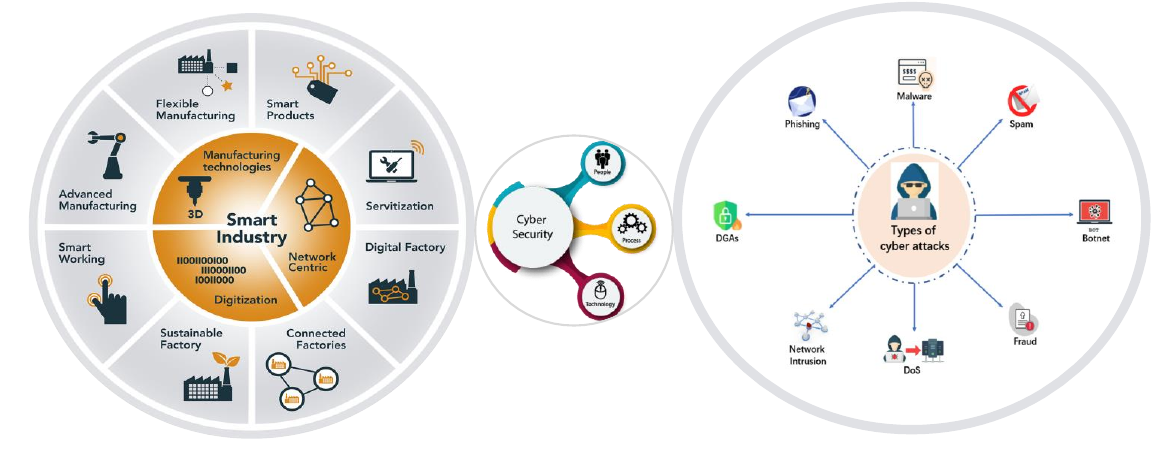}
\caption{Evolution of Industry 5.0: A visual representation highlighting the trajectory of smart industry development underscored by the imperative for robust cybersecurity measures in the face of escalating cyber attacks.} 
\label{Fig: Industry}
\end{figure*}

In Industry 5.0, we're witnessing higher reliance on IoT devices which makes systems more prone to cyber threats, and could results into serious damage and financial losses. To address the rising challenges of cybersecurity in the forthcoming Industry revolution, where smart cities and industries are evolving, various advanced security measures have been implemented. These include Security Information and Event Management (SIEM) systems, vulnerability assessment solutions, Intrusion Detection Systems (IDS), and user behavior analytics \cite{kiran2023intrusion,salam2023deep}. In this study, our focus is to evaluate the advancements in IDS security measures within this realm and to highlight the challenges that still need to be addressed. The automatic and real-time analysis of events within a computer system or network to indicate potential software or hardware security problems is achieved through IDS. Equipping communication systems with Intrusion Detection Systems (IDS) for monitoring operations, gathering alert intelligence, and mitigating threats and attacks poses a formidable challenge for intruders \cite{bace2001intrusion, czeczot2023ai}. However, as smart industries evolve and become more interconnected, as shown in Figure \ref{Fig: Industry}, cyber attackers are continuously probing networks to exploit access barriers and develop sophisticated cyberattack methods for their benefit.

The adaptation of ML and Deep Learning (DL) algorithms in IDS introduced various intelligent systems, which optimized the detection rate to a very high extent. The adoption of these ML/DL-based IDSs is because these techniques are more robust, accurate, and extensible as compared to other traditional detection techniques like rule-based, signature-based, and anomaly-based detection \citep{hussain2020machine, chou2021survey}. In general, the core foundation behind these complex algorithms is the mathematical and statistical concepts, which primarily perform pattern discovery, correlations or dependence and disparity of structured data, and represent their outputs in terms of probabilities and confidence intervals \cite{alazab2021deep, markevych2023review}. The primary types in ML include \emph{Supervised, Semi-supervised, Unsupervised, Reinforcement, and Active learning} techniques. Supervised learning is widely used when labeled data is abundant, while semi-supervised learning is suitable when labeled data is limited. Unsupervised learning is effective for exploring data structures and detecting anomalies. Reinforcement learning is utilized in decision-making scenarios with a reward system, and active learning aids in efficient data labeling \cite{sowmya2023comprehensive}. All these techniques play a key role in the foundation of DL algorithms, which are now trying to be adopted in the provision of intelligent Security Information and Event Management (SIEM) systems, vulnerability assessment solutions, Intrusion Detection Systems (IDS), user behavior analytic, etc. Besides the effectiveness of these intelligent AI-based modules, the \emph{transparency} of these opaque/black-box models and the \emph{justification} for their prediction is still a mystery. Such a lack of insight into the inner decision-making system of these opaque AI models raises trust issues in adopting these modules in the \emph{Industry 5.0} revolution \cite{sauka2022adversarial}.

\subsection{Scope of the survey}\label{subsec1}

This survey focuses on highlighting the critical challenges the security practitioners are being confronted with (i.e., the integration of successful security and defense measures in high-risk cyber-physical systems) in Industry 5.0 applications. Figure \ref{Fig: Industry} provides a visual representation of evaluation of industry 5.0 highlighting the trajectory of smart industry development underscored by the imperative for robust cybersecurity measures in the face of escalating cyber attacks \cite{koay2023machine,jeffrey2023review, aloqaily2022special}. This heightens the scope of the paper and the need to address this issue by critically analyzing the research trends in cybersecurity for smart industries. The paper mainly focuses on exploring the impact of explainability and interpretability concepts on cybersecurity practices.

\subsection{Related Surveys}\label{subsec2}
The continuous evolution of industrial paradigms has introduced transformative goals, emphasizing the creation of a resource-efficient and intelligent society. This trajectory seeks to elevate living standards and mitigate economic disparities through the integration of a hyper-connected, automated, data-driven industrial ecosystem \cite{taj2022towards, maddikunta2022industry, bobek2023industry}. This digital transformation promises to significantly enhance productivity and efficiency across the entire production process. These milestones become possible through the integration of AI/ Generative AI as a collaborative landscape, fostering innovation, optimizing resource utilization, and driving economic growth in smart industries \cite{rane2023chatgpt}. However, it is imperative to acknowledge that such advancements expose the system to an elevated risk of sophisticated cyber-attacks \cite{lechachenkocybersecurity}. The connected devices and networks of the autonomous industry infrastructure are more prone to hijacking, malfunctioning, and resource misuse threats, which necessitates extra security layers to safeguard from such threats. The conventional deployed security measures, that are, AI-based cybersecurity systems are still in progress to mature, and the need for developing robust and trustworthy security systems has become a trending goal for defenders to achieve \cite{ahmad2024communications, bhattacharya2022internet}.

In this essence, the need to comprehend and interpret the causal understanding of inferences made by AI-based learning models directed the attention of the research community toward the XAI research field. In the literature, the taxonomy transition of XAI has been evaluated based on trust building for human-machine interaction \cite{yang2023survey}. Due to the multi-disciplinary application of AI, different domains have explored the explainability and interpretability concepts with a broad concept. Summarized in Table \ref{tab:TABLE I}, for instance, In \citep{minh2022explainable, buijsman2022defining, dwivedi2023explainable}, the authors comprehensively review the origin of different XAI concepts and mechanisms. Due to the critical role of cybersecurity in the smart industry evolution, there has been a significant increase in research in finding reliable security and incident response mechanisms. Apart from physical security mechanisms, cyber-physical systems are prone to various cyber attacks, which are also comprehensively analyzed in the literature \citep{beg2023review, habib2023xai}. 

The adoption of explainability techniques in cybersecurity, specifically in Intrusion detection and prevention systems was thoroughly reviewed in the latest surveys. For instance, in \citep{chandre2023explainable, moustafa2023explainable}, the authors provide an overview of different XAI mechanisms adopted in intrusion detection systems. The recent advancement of autonomous transportation, smart cities, and automatic energy management and control systems are highly vulnerable to attacks, therefore most of the recent literature focusing on these areas \citep{ahmad2022developing, nwakanma2023explainable, castrocomprehensive}. Besides to get the causal understanding of the learning model, XAI has been adopted in exploiting ML intelligence after getting insights of the model. In this survey, we analyze the impact of explainability concepts on cybersecurity practices. We also emphasized the current trend of the Adversarial Explainable IDS (Adv-XIDS) concept, which is a major issue now explainable AI-based decision models confronting in cybersecurity.
 
\subsection{Contributions}\label{subsec3}
Based on serious threat vectors and their implications, in this paper, we analyze the adoption of different XAI methods in IDSs and examine the impact of interpretability on Cybersecurity practices in the Industry 5.0 applications. In detail, we provide an overview of the literature on XAI-based cybersecurity solutions for Industry 5.0 applications with a particular focus on existing solutions, associated challenges, and future research directions to overcome these challenges. To make it self-contained, we also provide an overview of the taxonomy of XAI. 
The main contributions of this paper are summarized as follows:
\begin{itemize}
	\item We provide a clear and comprehensive taxonomy of XAI systems.
    \item We provide a detailed overview of current state-of-the-art IDS, their limitations, and the deployment of XAI approaches in IDSs.
    \item We also discuss the exploitation of XAI methods for launching more advanced adversarial attacks on IDS.
    \item We also highlight the current cybersecurity challenges and potential solutions to ensure the safety and security of industry 5.0 applications.
\end{itemize}

The rest of the paper is organized as follows. Section \ref{sec:Methodology} presents the methodology adopted for conducting this survey by briefly describing the objective questions of this survey. Section \ref{sec:eXplainableAI} provides an overview of the eXplainable AI taxonomies. Section \ref{sec:Cybersecurity Challenges in Industry 5.0}, presents cybersecurity challenges in Industry 5.0. Section \ref{sec:Intrusion Detection Systems for Cybersecurity in Industry 5.0} presents conventional IDS and the evalutions of the systems from AI-based IDS to XAI-based IDS. This section also cover different type of explainability mechanisms, specifically Self-model, Pre-model and Post-modeling explainability techniques. Section \ref{sec:adversarialIDS} presents adversarial XAI techniques in cybersecurity with the focus on exploring the exploitation of explainability mechanisms for different adversarial attacks. In Section \ref{X-IDS Challenges and Future Research Directions}, we discuss the challenges in the current XAI-based IDS systems and future research opportunities. Finally, Section \ref{sec:conclusions} concludes this survey.

\begin{table}[htbp]
\centering
\caption{\textbf{Related Survey}}
\label{tab:TABLE I}
\begin{tabular}{|>{\centering\arraybackslash}p{0.6cm}|>{\centering\arraybackslash}p{1.8cm}|>{\centering\arraybackslash}p{1.5cm}|>{\centering\arraybackslash}p{1.3cm}|p{6.0cm}|}
\hline
\textbf{\fontsize{2}{2}\selectfont Ref.} & \textbf{\fontsize{1}{1}\selectfont XAI-Taxonomy} & \textbf{\fontsize{2}{4}\selectfont Cybersecurity} & \textbf{\fontsize{1}{1}\selectfont Adv-XIDS} & \textbf{\fontsize{9}{9}\selectfont Summary} \\ 
\hline
\cite{taj2022towards} \cite{maddikunta2022industry} \cite{bobek2023industry} & \ding{51} & \ding{51} & \ding{55} & These papers examine the role of explainability in AI systems and industrial processes, highlighting the potential of Industry 5.0 through advancements like big data processing, AI, drones, cybersecurity, robotics, additive manufacturing, and IoT.\\
\hline
\cite{yang2023survey} & \ding{51} & \ding{51} & \ding{55} & The survey discusses the applications of explainability in healthcare, finance, law, cybersecurity, education, and engineering sciences, studying each domain with different case studies.\\
\hline
\cite{minh2022explainable} \cite{buijsman2022defining} \cite{dwivedi2023explainable} & \ding{51} & \ding{55} & \ding{55} & The survey provides a study of XAI methods in Pre-model, interpretable model, and post-model level explainability.\\
\hline
\cite{buijsman2022defining} & \ding{51} & \ding{55} & \ding{55} & The paper defines explanation in XAI as answering what-if-things-had-been-different questions, emphasizing contrastive formats and generalizations.\\
\hline
\textbf{Ours} & \ding{51} & \ding{51} & \ding{51} & This paper delves into cybersecurity in Industry 5.0, focusing on Explainable AI-based Intrusion Detection Systems. We explore the taxonomy of explainability, address cybersecurity challenges, provide insights into ML-based IDS enhanced by XAI, and highlight how explainability can be exploited for adversarial attacks in IDS, demonstrating its dual nature in cybersecurity.\\
\hline
\end{tabular}
\end{table}

\section{\textbf{Methodology}}
\label{sec:Methodology}
In the context of Industry 5.0, our focus was on investigating X-IDS to explore cybersecurity solutions and challenges in the forthcoming industry revolution. Our objective was to analyze the array of approaches and techniques, particularly those utilizing big data and advanced analytics, to bolster security outcomes. This endeavor involved a thorough review of existing research and developments in the cybersecurity domain, specifically targeting Intrusion Detection and Prevention Systems (IDPS). Our research methodology included a systematic examination of academic papers, industry reports, and relevant literature from various sources to identify key trends, methodologies, and emerging practices related to Explainable IDS. We also critically evaluated these methodologies to determine their effectiveness in providing transparency and comprehensibility within the secure Industry 5.0 framework with additional evaluation of the threats posed by adopting these mechanisms. The synthesis of this extensive review serves as the foundation for our analysis and findings, contributing valuable insights to the field of cybersecurity and the pursuit of more transparent and interpretable IDS solutions. Our research questions are: \\ \par

\begin{enumerate}[label={\textbf{Q\arabic*:}},leftmargin=1.2cm]
    \item What are the key cybersecurity challenges in Industry 5.0 and how does IDS help in mitigating the impact of cybersecurity threats?
    \item Why do we need interpretable and explainable AI-based Intrusion Detection Systems (X-IDS) in the forthcoming industry 5.0 revolution?
    \item What are the primary techniques and methods used in X-IDS to enhance transparency and interpretability?
    \item What are the primary challenges and limitations associated with X-IDS?
    \item What are the security implications of adversaries gaining insights into the internal decision mechanisms of X-IDS systems, and how can these systems be safeguarded against potential exploitation?
        \item What are the emerging trends and future research directions in the field of X-IDS?
\end{enumerate}  
\par 

According to the questions raised in this review, the research was based on searching specific keywords and terms to find relevant papers that could help in answering these questions, and cover the existing state-of-art "\textit{Interpretability and Explainability in IDSs}" approaches that deal with the cybersecurity issues in the context of the forthcoming industrial revolution. Our goal was to focus on searching the most significant keywords including Artificial Intelligence (AI), Explainable Artificial Intelligence (XAI), Explainability, Black-Box, Industry 5.0, Cybersecurity, Intrusion Detection Systems (IDS), X-IDS, Adversarial XAI, in the top indexed scientific databases. Each study has been critically analyzed for the inclusion or exclusion of coverage of at least one of the research questions we based our review on.

\section{\textbf{Explainable AI Taxonomies}}
\label{sec:eXplainableAI}
The field of intelligent AI-based learning methods has evolved significantly, reaching a stage where a substantial portion of critical decisions relies on predictions from trained models. However, there exists a realm of intelligence where machines must justify their decisions in response to questions like "Why," "What," or "How." In simple words, a decision model should explain their decision in such a way that it could be acceptable with no dough, understandable with no difficulty and could be reliable to enhance trust between users and technology \cite{ holzinger2022explainable, speith2022review}. \par

This pursuit of interpretability and explainability goal in AI coined the term, eXplainable AI (XAI), in the research community, which embodies the idea of developing understandable AI models that are consistent with expert knowledge. The intuition behind XAI is rooted in the concept that humans should be able to comprehend and trust the outputs and recommendations provided by AI systems. XAI aims to bridge the gap between the inherent complexity of AI algorithms and human understanding by providing transparent and interpretable explanations for the AI model's outputs \cite{sauka2022adversarial, ferreira2023recommender}. A presentation for model's decision in a textual (\emph{Natural Language Explanation}) or visual artifactory (\emph{Saliency Map}) that provide easy understanding of the relationship between the input instance's variables and the model's output. Such explanations empower users to evaluate the trustworthiness, reliability, and fairness of AI systems, fostering informed decision-making and facilitating collaboration between humans and machines \cite{10.1145/2939672.2939778}. \par  

In the current landscape of machine learning (ML), there has been a renewed interest in explicating the decision processes of knowledge-based expert systems. This resurgence is driven by the imperative for intelligent ML models to provide not only accurate predictions but also transparent insights into their decision-making rationale. This shift has led to the development of diverse ethical and responsible AI approaches, as well as transparent and verifiable/decomposable machine learning techniques. The burgeoning field of eXplainable AI (XAI) has experienced exponential growth in research, witnessing the adoption of various approaches in diverse industrial applications, including autonomous systems, security and privacy management, healthcare, industrial engineering, finance, and smart agriculture. The suitability of these approaches is contingent upon the specific AI model, task requirements, and application context \citep{ ahmed2022artificial, le2023exploring}.\par

\begin{figure*}[ht]
  \includegraphics[width=\linewidth]{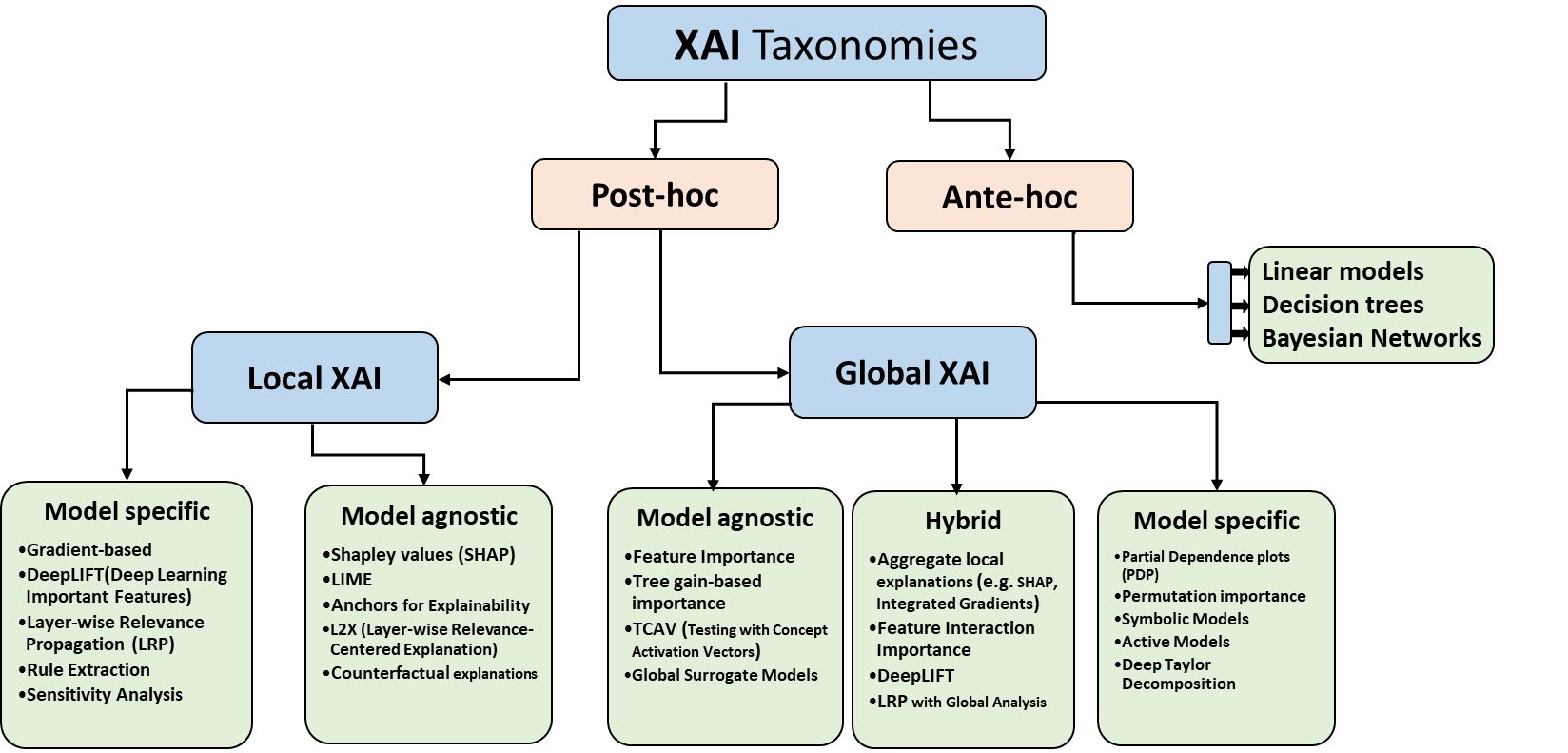}
  \caption{XAI Taxonomies}
  \label{fig:XAI Taxonomies}
\end{figure*}

In response to the inherent trade-off between ML performance and explainability, the research community has introduced two key XAI concepts: "Ante-hoc" and "Post-hoc" explainability methods \cite{gunning2019darpa}. Illustrated in Figure \ref{fig:XAI Taxonomies}, this discussion will briefly explore the taxonomy of XAI in the domain of security, with a specific focus on XAI-based Intrusion Detection Systems (X-IDS). This exploration involves a comparative analysis of the advantages and disadvantages inherent in popular proposed approaches. \par

\subsection{\textbf{Ante-hoc explainability}}
These are a kind of models that are self-explaining or can be called interpretable by design. Within this kind of transparency, these models could be interpreted in three levels including the model’s algorithmic level transparency, parametric level decomposability, and functioning level simulatability. Typical examples are linear regression, logistic regression, decision trees, random forests, Naïve Bayes and fuzzy inference, rule-based learning \cite{sarkar2022framework, BARREDOARRIETA202082}. The following examples very well meet the criteria of decision explainability, but these models perform very poorly on high-dimensional data. \par
From the Linear Models, for example, in \textit{Linear Regression}, the prediction is simply the weighted sum of input features. The weighted sum can be utilized as a measure of explainability because the model’s predicted target shows the linear relationship among the features. Additionally, the statistical measures associated with the linear regression model, such as p-values and confidence intervals, provide information about the significance and uncertainty of the coefficients. These measures can help assess the reliability and robustness of the model's explanations. Another type of regression analysis, i.e., \textit{logistic regression}, represents the target prediction as an estimated probability. This probability function is the magnitude of the coefficients that interprets an intrinsic sense of how much a feature is driving a model's prediction. Interpreting the coefficients in logistic regression involves examining their sign, magnitude, and statistical significance. Positive coefficients indicate a positive influence on the probability of the positive class, while negative coefficients suggest a negative influence. The magnitude of the coefficients signifies the strength of the relationship between the input feature and the probability. \cite{islam2021explainable}. \par
The \textit{Decision tree-based models} use facts and values rather than intuitions and are widely used for estimating the feature importance of linear and non-linear models. These models make predictions by partitioning the input features based on a series of decision rules, resulting in a tree-like structure. The path from the root to the leaf node interprets how the decision has been taken. The complexity comes with the increase in the number of nodes, also it cannot express the linear relationship between input and output. Very sensitive to a slight change in input data which makes it more challenging to interpret the output \cite{hanif2021survey}. \par
In the same way, \textit{Random Forest} and \textit{Gradient Boosting} are also part of the category of shallow learning models as they are built over decision trees in the form of an ensemble of decision tree models. These models are not enough simple to be interpreted, as the ensemble is not locally explainable due to the independent decision path of each tree, thus, only the global explanation, showing the general importance of the features in the target prediction, is obtained.  \par

Another intrinsically explainable method is, using \textit{Bayesian Networks} (BN) to learn the dependencies within a set of random variables through constructing an approximated Directed Acyclic Graphs (DAG). The graphical representation of a Bayesian network allows users to visualize the relationships and dependencies among variables. The structure of the graph provides insights into the cause-effect relationships and influences between variables. From the dependency graph, one can derive interpretation for the variables that are dominant in our system and can also gain insights into the probabilistic relationships between variables, understand the model's reasoning process, explain predictions or beliefs, assess the impact of evidence, and quantify uncertainties. These explainability mechanisms make Bayesian networks valuable for decision-making, risk analysis, and understanding complex systems with uncertain information. \par
Other self-descriptive models include \textit{Rule-based} learning models, which easily interpret their target prediction by using “if-then-else” rules. The descriptions, in the form of extensive background knowledge, are incorporated in the model with predefined \textit{set of rules and facts}. This technique provides a generic explainability of the diploid model but the deterministic nature and hard-to-scale for complex models make it difficult to be utilized. However, instead of defining specific rules for some tasks, such as IDS, deploying rule-based learning techniques can overcome the scalability issue. In rule-based learning, the goal is to find a set of if-then rules that accurately represent the patterns or associations present in the data. Rule learning techniques offer key advantages, such as interpretability, transparency, and human-understandable representations of patterns in the data. The discovered rules can provide insights into the relationships between features and outcomes, allowing for explanations and decision-making based on explicit rules.\cite{liu2021review}.\par

The above-discussed intrinsic models and the proposed interpretability techniques are pure transparent approaches and have achieved competitive performance in many regression and classification problems, but, these interpretability and explainability methods are limited to the model families comprising lower complexity. Some of the key constraints, such as model size, sparsity, and monotonicity can be considered, are the main reasons for the trade-off between performance and transparency. Thus, the complex model families, such as Ensembles, Artificial Neural Networks (ANN), Deep Neural Networks (DNN), and Support Vector Machines (SVM), which are considered opaque for not revealing the logic behind their predictions, are dealt with \textit{post-hoc} methods. Another recent explainable AI methods category is called \textit{Hybrid} approaches, which combine multiple explainable AI techniques to get appropriate interpretability with a high predictive performance \cite{dovsilovic2018explainable}.

\subsection{\textbf{Post-hoc explainability}}
\label{Post-hoc explainability}
The discernible reality is that many complex black-box models boast formidable predictive capabilities at the expense of limited explainability regarding their decision-making processes. In the realm of intelligent systems, accuracy, and interpretability stand out as primary characteristics. Addressing the imperative for transparency, surrogate explainers are essential, necessitating their development and application to interpret the rationale behind decisions made by sophisticated models.\cite{BARREDOARRIETA202082}. A prevalent and contemporary research avenue actively tackling the opacity challenge of complex black-box model families involves the utilization of post hoc explainability methods.

To elucidate the decision-making process of a trained model for a given input, post-hoc explainability addresses two overarching types of explanations: \textbf{Local explanations} and \textbf{Global explanations}. Local explanations are geared towards explaining how the model predicts outcomes for a specific instance, providing insights into the influence of a particular individual instance on a specific class. This granularity permits users to scrutinize the model's decision-making process at a detailed level \cite{van2023explainable}. Conversely, global explanations aim to assess the impact of all input features on the model's overall output. Through global explanations, one can comprehend how a model learns by manipulating the magnitude of specific features. This method is focused on understanding the model's general tendencies, feature importance, and decision boundaries. Both Local and Global explanations are further classified into \textit{model-agnostic} techniques, designed for post-hoc explainability applicable to AI-based models of any kind, and \textit{model-specific} techniques tailored for specific models \cite{vale2022explainable}.  \par

\textbf{\textit{Model-agnostic}} posthoc explainability techniques are designed with the concept of mimicking their complex structured antecedent with reduced complexity and visualizing their behavior understandably. These techniques directly focus on the model’s predictions rather than focusing on its internal representations. The potential to be deployed into any learning model, i.e., independent of the internal logic, the research community is mostly focusing on these techniques \cite{speith2022review}. Most of the Post-hoc model-agnostic explanation techniques are based on the concepts of quantifying the influence of each feature on the model’s prediction by a principle of simulating feature removal. This kind of explanation has been named \textit{removal-based explanations}, where different feature-removing techniques are used to examine the impact of the absence of a feature on the prediction \cite{covert2021explaining}. Other types of explanations for model-agnostic techniques are based on model simplification, feature relevance estimation, and visualization techniques.\par 

The most widely used model-agnostic explainability approaches include LIME (Local Interpretable Model-agnostic Explanations) and SHAP (SHapley Additive exPlanations). The LIME technique is learning a local model parallel with their antecedent opaque model to linearly explain its prediction. The key idea behind LIME is to approximate the complex decision boundary of the original model in the local neighborhood of a specific instance using a simpler and interpretable surrogate model \cite{confalonieri2021historical}. By focusing on the local context, LIME provides explanations that are more interpretable and relevant to the specific instance of interest. The local explainability is generated by the surrogate model's coefficients or feature importance measures, which are used to highlight the features that have the most significant impact on the prediction for that specific instance. LIME offers several techniques to provide local explanations for individual predictions across different data domains. Text-based LIME focuses on generating local and global interpretable explanations for text classification models by leveraging bag-of-words or word embeddings \cite{alantari2022empirical}. Image-based LIME perturbs image pixels to approximate local regions contributing to image classification predictions, and providing visual explanations \cite{lin2021towards}. Tabular LIME is designed for structured tabular data and employs locally interpretable models, such as linear regression or decision trees to approximate model behavior. Time Series LIME handles time-dependent features in time series data and provides explanations for time series prediction models. Concept-based LIME explains predictions based on higher-level concepts or groups of features, offering insights into the reasoning behind model predictions. These techniques demonstrate the versatility of LIME in delivering meaningful and interpretable explanations for diverse ML applications and data types \cite{ 10.1145/2939672.2939778}. \par
Shapley additive exPlanations (SHAP) technique, which is a game theory-based algorithm assigning each feature an importance value (SHAP value). This value indicates the average expected marginal contribution of a single feature to the prediction  to explain that specific model prediction. SHAP is a model-agnostic technique that explains the predictions of machine learning models \cite{NIPS2017_8a20a862}. It works by calculating Shapley values, which measure the contribution of each feature to the final prediction. SHAP considers all possible coalitions of features and evaluates their contributions based on how their inclusion or exclusion affects the prediction compared to a baseline reference value. By calculating the average contribution of each feature across different coalitions, SHAP assigns Shapley values to each feature, indicating its impact on the prediction. These Shapley values are then used to generate explanations for individual predictions, highlighting each feature's relative importance and direction of influence \cite{oseni2022explainable}. The result is a comprehensive and interpretable understanding of a model's predictions, capturing both the individual and interactive effects of features. There are various techniques within the SHAP framework, such as Kernel SHAP, Tree SHAP, Deep SHAP, and Linear SHAP, which are tailored to different types of models. Kernel SHAP applies a sampling-based approach to estimate Shapley values, while Tree SHAP adapts the concept to tree-based models. Deep SHAP utilizes a deep learning reformulation for complex neural networks, and Linear SHAP simplifies the calculation for complex linear models \cite{alenezi2021explainability}. \par

Other prominent approaches include Local and global visualizations of model predictions such as Accumulated Local Effect (ALE) plots, one/two-dimensional Partial Dependence Plots (PDPs), Individual Conditional Expectation (ICE) plots, and decision tree surrogate models. ALE, PDP, and ICE plots are all popular techniques in the field of explainability. ALE plots provide insights into the average change in predictions as specific features vary, considering the effects of other variables. They capture the cumulative effects of feature changes on model predictions. PDP plots, on the other hand, showcase the relationship between one or two features and the model's predictions while controlling for other variables \cite{speith2022review}. They reveal the direction and magnitude of feature effects and can uncover nonlinearities and interactions. ICE plots take a similar approach to PDP plots but provide a more granular view by displaying the predictions for individual instances as the feature(s) of interest vary. These plots offer a more personalized understanding of feature effects. All three techniques are model-agnostic and help in interpreting and explaining the behavior of machine learning models, aiding in feature selection, decision-making, and model evaluation. In the same way, Decision tree surrogate models serve as interpretable approximations of complex models, allowing for a more understandable representation of their decision-making process \cite{vale2022explainable}. \par 

\textbf{\textit{Model-specific}} posthoc explainability covers those interpretable models that show transparency by design but due to their complex internal decision structure, it cannot be easily interpreted like Ante-hoc explainability models. These complex models need additional layers by leveraging the specific structural and behavioral knowledge of a particular type of model to provide interpretability. In this type of interpretability, the most prominent utilized ML models achieving higher accuracy at the expense of explainability are \textit{Tree ensembles}, \textit{Random Forests}, \textit{Gradient Boosting}, \textit{Linear Regression}, \textit{Bayesian Networks} and \textit{Support Vector Machine (SVM)} models. These are the extended form of shallow learning models, which adopt complexity in their internals to be utilized for large complex decision problems, also, the explainability methods for each learning model vary according to each target model type \cite{belle2021principles}. \par 
The idea of ensembles came as a concept to circumvent the overfitting issue by generalizing the single tree prediction/regression into an aggregated prediction of different combined decision trees. Examples include Tree ensembles, Random Forests, Gradient Boosting, and Multiple Classifier Systems, where the final decision is the combination of the aggregated base classifier’s decision. The state-of-the-art techniques utilized for explaining the internal decision mechanism of ensemble models are based on \textit{explanation by simplification} and \textit{feature relevance} techniques \cite{kim2022lightweight}. A collective behavioral explanation of the ensemble models generated by various simplification techniques, includes weighted averaging (Where weights reflect the importance of each model's contribution, providing a simplified explanation), Model Distillation (The Distilled model provides a simplified explanation by mimicking the complex ensemble behavior into a simple interpretable form) \citep{tan2018distill, liu2021improved}, G-REX (Genetic-Rule Extraction) is another approach for rule extraction from complex black-box models where a set of rules from the ensemble capture the collective behavior \cite{johansson2007inconsistency}), Feature Importance analysis (Identifying the most influential features using permutation importance or information gain to rank the importance and provide a simplified explanation). Explanation by these kinds of techniques provides insights into the collective behavior of the ensemble models and provides explanations for their predictions \cite{ konig2008g}. \par

Support Vector Machine (SVM), which is more complex in structure than ensembles, because they construct high-dimensional hyper-planes to find a good separation between two class instances. The explainability for this kind of learning model is also based on explanation by \textit{model simplification, Counterfactual, and example-based explanations}. Model simplification refers to the process of interpreting and understanding the decision boundaries (support vectors) and the importance of features in SVM. The features associated with the support vectors present their importance and provide a more interpretable representation of the SVM model. Counterfactual explanations aim to provide insights into how the SVM model's decision would change if certain features or inputs were modified. These explanations present hypothetical scenarios that demonstrate the model's sensitivity to changes in input variables. By identifying the minimal changes required to alter the SVM's decision, counterfactual explanations offer valuable insights into the decision-making system \cite{moustafa2023explainable}. In the same way, Example-based explanations involve using representative examples from the dataset to illustrate the SVM model's decision process. These examples can showcase how the model detects different instances and highlight the salient features that influence the detection process. By examining multiple examples, patterns, and relationships between features and decisions can be discerned, leading to a better understanding of the model's behavior.  Some of the techniques to explain the decision of the SVM also include rule-based methods where the intention is to extract rules from a trained SVM classifier. Another concept of explaining the SVM model is based on a color-based nomogram (approximating graphical computation of mathematical function) \cite{ van2016explaining}. \par
Explainability of \textit{deep learning models} is more important due to their applications in a diversified set of domains including Industry 5.0. The deep learning models include Multi-Layer Perceptron (MLP), Convolutional Neural Networks (CNN), and Recurrent Neural Networks (RNN). Due to their complex black-box nature, most of the explainability techniques are post hoc, which are mostly based on model simplification, feature importance estimators, explanation in text or visualizing saliency maps, collection of local and global explanations \cite{oseni2022explainable}. Some recent works adopted \textit{Hybrid} approaches, which are based on background knowledge of the model’s logical constraints. These techniques show robust performance for deep ML models as they are based on the combination of the expert’s written rules and the knowledge generated from the decision tree algorithm \cite{szczepanski2020achieving}. \par
After briefly discussing the XAI taxonomies, now we are heading to give an extensive overview of the demand for explainability in AI-based cybersecurity applications, especially explainability in IDSs. This is a very critical and mature research area progressed into very intelligent IDS systems and due to the integration of this paradigm as a crucial part of Industry 5.0, the need for the IDS system’s transparency become a pressing matter.

\section{Cybersecurity Challenges in Industry 5.0}
\label{sec:Cybersecurity Challenges in Industry 5.0}
Industry 5.0, which is mainly focused on the collaboration of humans and machines for different tasks, involves the integration of several technologies, such as AI, data analytics, IoT, augmented and virtual reality, and improved man-machine interfaces (MMI) allowing workers to carry out different operations. This enhanced interconnectivity exposes smart industries to a diversified set of cybersecurity challenges and threats, which may lead to a disastrous operational environment, putting workers at risk and halting production. Some of the most common cybersecurity threats to Industry 5.0 include:

\begin{itemize}

    \item \textbf{Expanded Attack Surface}: The enhanced interconnectivity has significantly increased the number of entry points for cyber attacks in Industry 5.0, making it more challenging timely detect and guard against different cyber attacks \cite{hassan2024systematic}. For instance, industries need to collect and analyze data for different tasks, such as customer behavior, optimization of marketing campaigns, better supply chain management, and predictive maintenance of the machines, etc. However, this data could be used by attackers for malicious activities. Thus, stricter access control and data management policies and techniques need to be incorporated to ensure the data is used for improvement purposes.  
    
    \item \textbf{Social Engineering}: 
    Social engineering, which exploits human mistakes/errors instead of technical vulnerabilities, is one of the major cybersecurity threats in the modern world. In recent years, social engineering tactics emerged as one of the most effective ways of obtaining sensitive information. Some of the common cybersecurity threats based on social engineering tactics include phishing, baiting, pretexting, malware, tailgating, and vishing. In Industry 5.0, due to the expanded human-machine collaboration, social engineering attacks have become a serious concern that needs attention \cite{rajabion2023industry}. 

    \item \textbf{Cloud Vulnerabilities}: Cloud computing, which aims at delivering remote computing services and storage services, such as data analytics and databases, is an integral part of several industry 5.0 applications \cite{maddikunta2022industry}. For instance, the technology can support industries with different types of manufacturing applications/tools, such as IoT-based real-time data access and monitoring and APIs for data normalization from diverse sources \cite{adel2022future}. However, cloud computing also brings several security challenges. For instance, the vulnerabilities of third-party software/applications, insecure APIs, and cloud data governance further expand the attack surface.

    \item \textbf{IoT Vulnerabilities}: IoTs are an integral part of Industry 5.0 enabling the industries to collect and analyze data on different aspects of the industry through connected devices and sensors. The data is then utilized in decision-making. However, it also brings several security challenges. For example, ensuring the security of a large number of 
IoT devices are very challenging, and inadequate protection may lead to different types of cyber attacks.

    \item \textbf{Supply Chain Vulnerabilities}: Industry 5.0 offers several benefits in supply chain management by extending the collaboration between humans and intelligent robots \cite{maddikunta2022industry}. However, the vulnerabilities, introduced by the complexities of the supply chain and the dependencies among partners, could expand the attack surface of cyber attacks in Industry 5.0.

    
\end{itemize}

The expanded attack surface in Industry 5.0 makes smart industries vulnerable to different types of cyber attacks. However, thanks to recent developments in AI and ML it is possible to monitor network traffic and identify unusual/suspicious activities, preventing cyber attackers from getting into the system. In the next section, we discuss these systems in more detail. 

\section{\textbf{Intrusion Detection Systems for Cybersecurity in Industry 5.0}}
\label{sec:Intrusion Detection Systems for Cybersecurity in Industry 5.0}

Cybercrimes are not limited to a specific region but are global threats of breaching defined access rules for an electronic system and their inflicted damages are increasing exponentially with the advent of new technologies. The damage can be unauthorized access to the system and making it unavailable to the authorized one, stealing confidential data or some ransomware, damaging the system’s functionality, or destroying data integrity. Nowadays, cybersecurity is considered an initial step for every starting setup where the communication devices are connected through the internet. An individual system could be made secure efficiently but whenever there is a need for connection to another remote system, the threat of the violation of security policies, that are, Confidentiality, Integrity, and Availability (CIA), becomes prone to occur \cite{article}. To circumvent these threats, numerous cybersecurity techniques have been proposed in the literature including Anti-Virus software, Firewalls, IDS, and Intrusion Prevention Systems (IPS). In this everlasting competition between the attackers and the defenders, now the systems are coming up with very intelligent security systems that sometimes outperform human-level intelligence \cite{9689765}. 

\begin{figure*}[ht]
  \includegraphics[width=\linewidth]{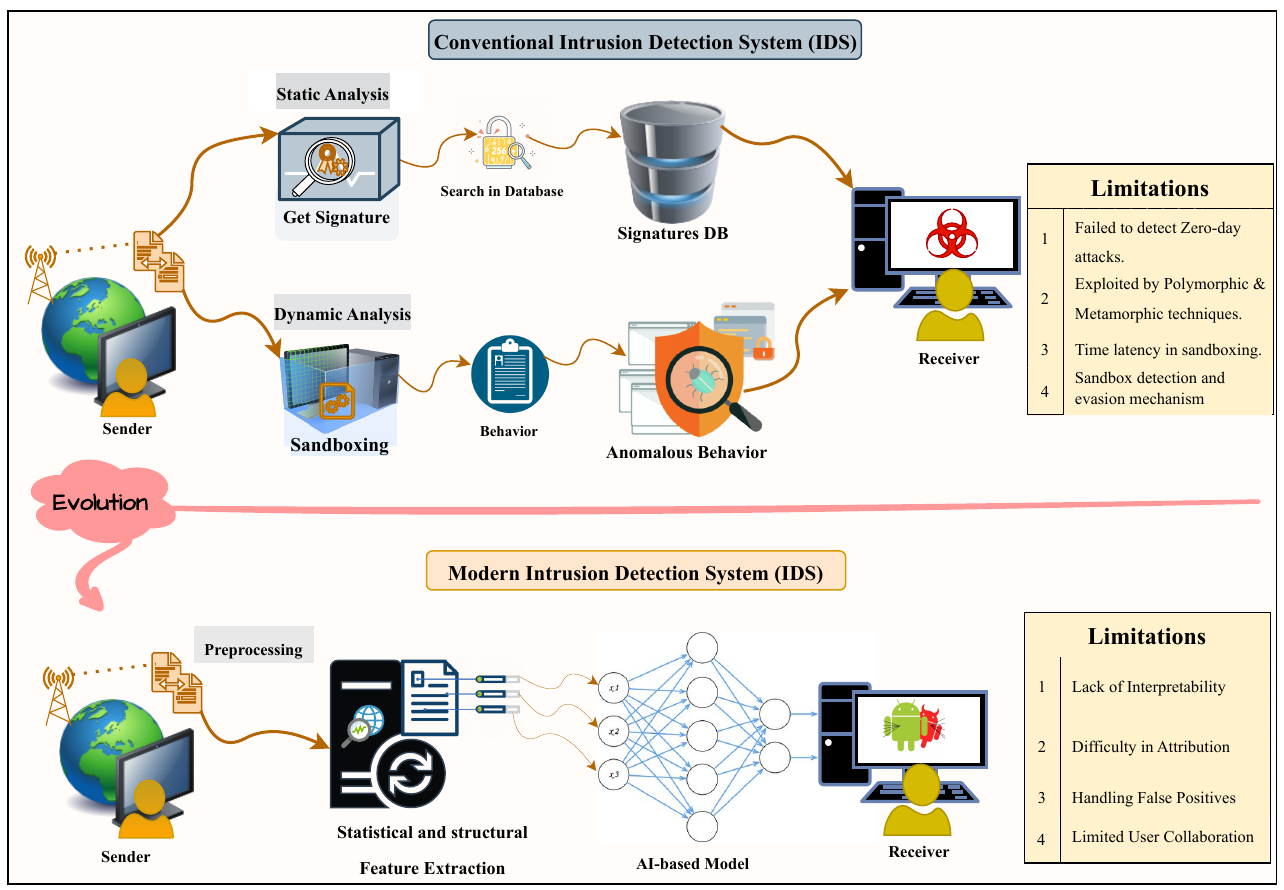}
  \caption{The evolution of conventional Intrusion Detection Systems (IDSs) to the Moderen IDSs.}
  \label{fig:IDS-Evalution}
\end{figure*}

Furthermore, in the context of Industry 5.0, the importance of cybersecurity is paramount. With the increasing interconnectedness of systems and the rise of automation, the risk of cyber-attacks targeting industrial systems has grown significantly. Cybersecurity in Industry 5.0 is not only about protecting data and systems from unauthorized access but also about ensuring the integrity, availability, and confidentiality of critical information and processes \cite{ahmad2024communications}. As Industry 5.0 relies heavily on interconnected devices and systems, any compromise in cybersecurity could lead to severe disruptions in industrial operations, financial losses, and potential safety hazards. Therefore, the development and implementation of advanced cybersecurity measures, including ML-based IDS, are essential to safeguard the integrity and security of Industry 5.0 systems \cite{kiruthika2024role, khan2024need}. In this section, we provide a detailed overview of conventional IDSs and their evolution to ML-based IDS by highlighting its key aspects with additional analysis of how XAI is playing its part in improving its effectiveness.

\subsection{\textbf{Intrusion Detection System (IDS)}}
\label{IDS}
In the literature, the conventional method for the IDS problem is tackled by Signature-based Intrusion Detection Systems (S-IDS) and Anomaly-based Intrusion Detection Systems (A-IDS). In S-IDS, the new pattern is just matched with the previously known attack patterns, also called Knowledge-based Detection \citep{liao2013intrusion, liu2019machine}. These techniques obey the idea of building a database of intrusion instances signatures and with this database, each new instance’s signature is matched, as shown in the upper part of Figure \ref{fig:IDS-Evalution}. This kind of detection technique failed to detect zero-day attacks, also the polymorphic and metamorphic techniques introduced in malware make it hard for IDS to identify the same malware with different shapes.\par 
The solution for the polymorphic and metamorphic problem has been addressed by an Anomaly-based Intrusion Detection System (A-IDS), where the variant is analyzed in a sandbox environment to analyze their behavior. Another technique for the analysis is creating a baseline of the normal behavior of the computer system using ML, statistical-based, or knowledge-based methods \cite{ahmad2021network}. After creating a decision model, any significant deviation between the observed behavior and the model’s characteristics is regarded as an anomaly and the model classifies it as an intrusion. From the traditional sandbox analysis perspective, A-IDS performs well in terms of detecting zero-day attacks as well as in detecting polymorphic and metamorphic properties of intruders. However, the issue is the detection speed, because S-IDS has a better speed of detection than A-IDS also the sandbox evasion mechanisms by attackers failed the conventional systems (Figure \ref{fig:IDS-Evalution}). The advent of powerful ML techniques has somehow overcome these issues by automatically discovering the essential difference between normal and abnormal data with high accuracy \citep{zhang2022explainable, kaur2023artificial}.\par 

ML solutions are based on the generalization of the given data to make a good prediction for the unknown data. These techniques perform well when we have sufficient training data. The performance of the ML-based IDS models depends on the quality of valuable information of the datatype, also the data acquisition should be easy, and fast, and will reflect the behavior of the source (i.e., the hosts or networks) \cite{8586840}. The common data sources for ML-based solutions include packets, function or action logs, sessions, and flow controls from connected devices. In the literature, the feature-based datasets, such as \textit{DARPA 1998, KDD99, NSL-KDD, CIC-IDS/DDoS-(2017-19)} and \textit{UNSW-NB15}, are used as benchmark datasets \citep{lippmann20001999, 5356528, 7348942}. \par


Now, let's discuss the use of multiple data types for detecting different attacks because each data type reflects a specific attack behavior, for example, the system functions and actions logs show the host’s behavior while the session and network flow reflect network behavior. Thus, according to the attack characteristics, appropriate data sources are required to be selected for the collection of valuable information \cite{ liu2019machine}. The header and application data present in the communication unit called a packet, contains details that could be used to detect U2R (User to root access) or R2L (Remote to Local access) attacks. The packet-based IDS include \textit{packet parsing-based} and Payload analysis-based detection. Another detection scheme is network flow-based attack detection, which is mostly used for DOS and Probe attacks. These methods include feature engineering-based and DL-based detection \cite{faraj2020taxonomy}. \par 
There are also some possible session creation-based attacks, which can be detected by using session statistical information datatype as an input vector for a decision model. Finding a sequence in the session’s packets can give detailed information about the session interaction, which is also targeted in the literature by using text processing technologies like CNN, RNN, and LSTM as an encoding method to extract spatial features from the session’s packets. There is another important attack detection method based on recorded logs by operating systems or application programs. The data contains system calls, system alerts, and access records. This kind of detection system needs to have an expert knowledge of cybersecurity to get an understanding of the recorded logs. The recent detection approaches include Hybrid methods which is the combination of rule-based detection and machine learning techniques. Other detection methods include text analysis techniques, where the system logs are considered plain text. A prominent method is the use of an n-gram algorithm to extract features from the text file and pass them to the classifier to perform detection and classification \cite{ liu2019machine}. \par

\subsection{\textbf{Explainable IDS (X-IDS)}}
\label{X-IDS}
\begin{figure*}[ht]
  \includegraphics[width=\linewidth]{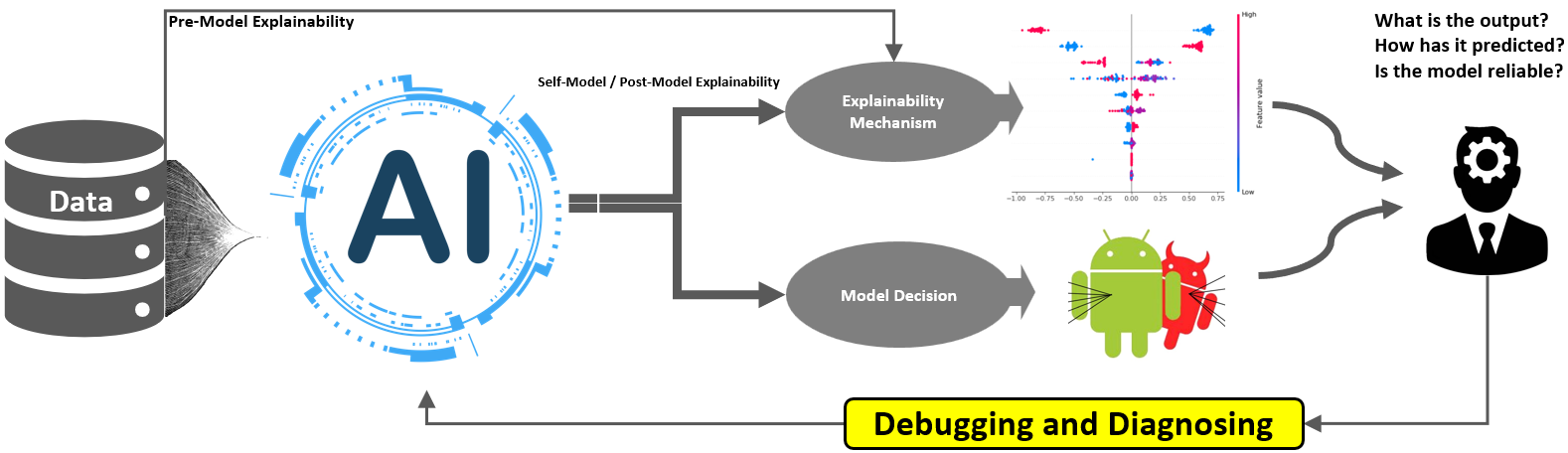}
  \caption{Explainable IDS (X-IDS), where a stakeholder tries to understand the decision taken by the model against an instance also with the intuition of debugging and diagnosing the model to enhance the performance. }
  \label{fig: Explainable AI}
\end{figure*}

From the above discussion, it becomes clear that most of the intelligent intrusion detection systems are based on complex ML techniques that perform very well in intrusion detection. However, another aspect of IDSs that needs to be considered in the design of such systems is transparency in the decision-making process. The system developer should know the answer to “Why” or “How” questions about the IDS model, to develop a more reliable, secure, and useful solution for the security problems (Figure \ref{fig: Explainable AI}). The explanation in the IDS system could be an explanation for alerts generated by IDS, also a reason for a decision made either anomaly or benign, also provide an indicator of compromise for a security analyst in operation center \citep{ szczepanski2020achieving, rjoub2023survey}. The need and usefulness of explanation in security systems were first proposed by Vigan et al \cite{vigano2020explainable}, by highlighting the need for explanations to better understand the core functionality of the intelligent system in terms of the "Six Ws" paradigm. They described the importance and enhancement of the security system by providing answers to these $Ws$, which are "Who", "What", "Where", "When", "Why" and "How". \par

\begin{sidewaystable}[htbp]
\caption{Explainable AI-Based Intrusion Detection Systems}\label{tab:TABLE II}
{\fontsize{6}{8}\selectfont
\begin{tabular}{|p{1cm}|p{2.5cm}|p{2.5cm}|p{2.5cm}|p{2.5cm}|p{2.5cm}|p{4.5cm}|} 
\hline
Ref. & Threat Addressed & IDS Datatype & Dataset & Detection Model & XAI Algorithm & Explanation \\
\hline

\cite{sinclair1999application} & Intrusion Detection & NIDS & Analyst-designed training sets from archived network events & Genetic algorithm, ID3 & Rule-based Explainability & Generating rules for classifying normal network connections from an anomalous one, based on expert-domain knowledge. \\
\hline

\cite{mahbooba2021explainable} & DOS, R2L, U2R, PROBING & NIDS & KDD-99 & ID3 algorithm & Rule-based Explainability & Generating rules by Rattle package in R and visualizing in exploratory plots. \\ 
\hline

\cite{manoj2023explainable}, \cite{10309785} & DOS, R2L, U2R, PROBING & NIDS & SCADA VM, N-BaIoT & DT, RF, GNB, SVM, K-NN & Rule-based Explainability & Generating rules by Tree nodes to visualize the decision-making process as exploratory plots. \\ 
\hline

\cite{yang2023bayesian} &  DOS, R2L, U2R, PROBING &  Host-based IDS (NIDS) & Tracer FIRE 9 (TF9) and Tracer FIRE 10 (TF10)   &   Bayesian Network (BNs) &  Rule-based Explainability & Visualizing network graph of the Bayes' Rule, where the relation of single feature to the target variable has been found via Conditional Probability Tables (CPTs). \\
\hline

\cite{zolanvari2021trust} &  Industrial IoT Security & IIoT-based IDS &  WUSTL-IIoT, NSL-KDD, UNSW-NB15  &  Artificial Neural Network (ANN) &  Transparency Relying Upon Statistical Theory (TRUST) system & Employing mutual information for ranking variables and select the most impactful ones on the ANN's outputs, naming them as representatives of the classes. \\
\hline

\cite{le2022classification} &  Industrial IoT Security & IoT-based IDS & IoTID20, NF-BoT-IoT-v2, NF-ToN-IoT-v2  & Ensemble tree (DT, RF) &  SHapley additive exPlanations (SHAP) & The ensemble model's outputs are plotted in the form of heatmap and decision plot using SHAP explanation techniques. \\
\hline

\cite{alani2022botstop} \cite{ALANI2023103409} \cite{Alani2022PAIREDAE} &  Android devices Malware detection &  Android Application-based IDS & MalMem-2022, Drebin-215, Malgenome-215, CICMalDroid2020  &  RF, LR, DT, GNB, XGB, SVM & SHAP & Feature importance for the selected features are calculated by Gini importance using node impurity, which are then plotted as an explanation summary for highest impact features. \\
\hline

\cite{alani2022towards} \cite{alani2023explainable} \cite{alani2023xrecon} & Industrial IoT Security & IoT-based IDS &  IoTID20, NF-BoT-IoT-v2, NF-ToN-IoT-v2  &  RF, LR, DT, GNB, XGB, SVM & SHAP & Feature importance for the selected features are calculated by Gini importance using node impurity, which are then plotted as an explanation summary for highest impact features. \\
\hline

\cite{gurbuz2023explainable} & Malicious Traffic Detection in IoT Healthcare Networks & IoT-based IDS)  & Intensive Care Unit(ICU) dataset  &  RF, DT & SHAP, LIME, ELI5, Integrated Gradients (IG) & Visualizing the contribution of each feature in the model's decision using Shapash Monitor explanation interface. \\
\hline

\cite{patil2022explainable} & Man-In-The-Middle (MITM), DoS,  Mirai botnet, Port/OS scanning, Host scanning & IoT-based IDS) & CICIDS-2017  & Voting Classifier & Local Interpretable Model-agnostic Explanation (LIME) & Plotting the contribution of each feature in the model's decision using LIME. \\
\hline

\cite{zebin2022explainable} & DNS over HTTPS (DoH) attacks & Network-based IDS & CIRA-CIC-DoHBrw-2020  &  Random Forest (RF) & SHAP & Highlighting the features which are contributing in the underlying decision of the model using SHAP values. \\
\hline

\cite{sivamohan2023optimized} & Glastopt, Dionaea, Cowrie, Canarytokens, DoS, R2L, U2R, Probe attacks & Network-based IDS (NIDS) & Honeypot and NSL-KDD datasets & Bidirectional Long Short-Term Memory (BiLSTM) & LIME, SHAP & Focuses on generating Global and local faithful explanations by approximating the behavior of the BiLSTM model around a specific instance of interest. \\
\hline

\cite{wang2023network} & DOS, R2L, U2R, PROBING & Network-based IDS (NIDS) &  KDD-99  & CNN-LSTM  & LIME, SHAP & LIME mechanism enables the model to interpret each individual factor and their impact on output. A decision tree generated from the top-most influential features which are then visualized using SHAP interpretation.  \\
\hline

\cite{tanuwidjaja2023hybrid} & DOS, R2L, U2R, PROBING & Network-based IDS (NIDS) & Ton-IOT Windows &  RF  & LIME , SHAP & Employed three primary techniques—variable importance plot, individual value plot, and partial dependence plot—to explain the decision-making process of the RF model.  \\
\hline

\cite{mills2019efficient} &  Malware Detection & File Content Analysis & VX Vault and Virus Share based generated dataset  & Random-Forest classifier & Visualizing Decision Tree & Presenting the trees that had classified a process as malware or benign and the relevant decision nodes.\\
\hline

\end{tabular}
}
\end{sidewaystable}

\begin{sidewaystable}[htbp]
\caption{Explainable AI-Based Intrusion Detection Systems - Continued}\label{tab:TABLE III}
{\fontsize{6}{8}\selectfont
\begin{tabular}{|p{1cm}|p{2.5cm}|p{2.5cm}|p{1.5cm}|p{2.5cm}|p{2.5cm}|p{4.5cm}|} 
\hline
Ref. & Threat Addressed & IDS Datatype & Dataset & Detection Model & XAI Algorithm & Explanation \\
\hline

\cite{ ables2022creating} \cite{ables2023explainable} & Web based, Brute force, DoS, DDoS, Infiltration, Heart-bleed, Bot and Scan &  Host-based and Network-based IDS &  NSL-KDD, CIC-IDS-2017 &  Population-base Self-Organizinng Maps (POPSOM) implementation  & Self Organizing Maps (SOMs) based X-IDS system  & Produce robust, explanatory visualizations of the SOM model and create accurate IDS predictions. \\
\hline

\cite{lundberg2022experimental} & DoS, ID fabrication &  In-Vehicle IDS (IV-IDS) &  Survival Analysis Dataset for automobile IDS &  Deep Neural Network (DNN)  & Visualization-based Explanation, (VisExp)  & A dual swarm plot is created to display normal Controller Area Network (CAN) traffic on top and intruder's traffic at the bottom based on SHAP-values distribution. \\
\hline

\cite{al2022xai} & Adware, Banking malware, SMS malware, Riskware, Brute Force FTP, DoS & File Content Analysis, Network-based IDS & MalDroid20, CIC-IDS2017 & Deep Neural Network (DNN) & DALEX framework & DALEX employs a permutation based algorithm to find the significance of individual variables which enhances the DNN prediction performance. \\
\hline
\cite{malik2022xai} & Adware, Banking malware, SMS malware, Riskware, Brute Force FTP, DoS & File Content Analysis, Network-based IDS & MalDroid20, CIC-IDS2017 & Deep Neural Network (DNN) & SHAP & Fine-tuning the DNN prediction performance through the combination of adversarial training and XAI. \\
\hline
\cite{lanfer2023leveraging} & Denial-of-Service (DoS), Probe attack types & Network-based IDS (NIDS) & NSL-KDD & Random Forest (RF) & SHAP & Utilized SHAP beeswarm plots to visualize the explanations of the target class individually. \\
\hline
\cite{lu2022does} & Android Malware detection & File Content Analysis & DREBIN & SVM, BERT & Feature Importance & Inspired by MPT, the method minimizes variance in prediction score changes and attribution values, signifying higher feature attribution for impactful changes in model predictions. \\
\hline
\cite{SARHAN2022100359} & Brute-force, Bot, DoS, DDoS, Infiltration, Web attacks & Network-based IDS (NIDS) & CSE-CIC-IDS2018, ToN-IoT, Bot-IoT & Multi Layer Perceptron (MLP), Random Forest (RF) & SHAP & Calculating Shapley values for features to calculate their contribution in the final decision and can also identify the key influencing features from the dataset. \\
\hline
\cite{oseni2022explainable} & DoS and DDoS in IoT/IoV networks & Network-based IDS (NIDS) & ToN\_IoT dataset & Deep Neural Network & Deep SHAP technique & Interpret and Combines the SHAP values calculated for smaller parts of a neural network into SHAP values for an entire network through the back-propagation of DeepLIFT’s multipliers. \\
\hline
\cite{alani2022deepiiot} \cite{alani2023arp} & Command injection, DoS, Reconnaissance, Backdoors & Industrial IoT-based IDS (IIoT-IDS) & WUSTL-IIOT-2021 & Deep neural network (DNN) & SHapley additive exPlanations (SHAP) & Through DeepExplainer, the SHAP values are used to provide insights into the decision-making process of DeepIIoT. \\
\hline
\cite{kalutharage2023explainable} & DDoS attacks on both Internet of Things (IoT) and traditional networks & IoT-based IDS (IoT-IDS) & USB-IDS dataset & Fully connected autoencoder model with RELU & Kernel SHAP & The autoencoder model uses SHAP values to identify which top-R features contributed to each of the significant reconstruction errors. \\
\hline
\cite{muna2023demystifying} & Industrial IoT Security & IoT-based IDS (IoT-IDS) & IoTID20 dataset & XG-Boost & LIME, TreeSHAP, ELI5 & Lime explains feature contributions, SHAP combines feature importance with effects, and ELI5 reveals feature weights in prediction. \\
\hline
\cite{abou2022novel} & IoT-network Security & IoT-based IDS (IoT-IDS) & UNSW-NB15 & DNN & RuleFit, SHAP & Calculating the feature importance values for the decision model. \\
\hline
\cite{Marino2018AnAA} \cite{da2023false} & DOS & Network-based IDS (NIDS) & NSL-KDD99, LYCOS-IDS2017 & Linear Model (LM), Multi Layer Perceptron (MLP) & SHAP, Adversarial machine learning & Generating visual explanations for incorrect estimations made by a trained model and identifying the features which are responsible for misclassification. \\
\hline
\cite{szczepanski2020achieving} & DDoS, XSS and SQL Injection attacks & Anomaly-based IDS (AIDS) & CICIDS2017 & ANN with Principal Component Analysis (PCA) & Decision Trees trained using microaggregation & Library dtreeviz generates a plot showing a tree’s structure, as the path leading to the prediction with highlighted important features. \\
\hline
\cite{wang2020explainable} & DOS, R2L, U2R & Network-based IDS (NIDS) & NSL-KDD & One-vs-All classifier, Multiclass classifier & SHAP & Combines the local and global explanations to improve the interpretation of IDSs. \\
\hline

\end{tabular}
}
\end{sidewaystable}

\begin{sidewaystable}[htbp]
\caption{Explainable AI-Based Intrusion Detection Systems - Continued}\label{tab:TABLE IV}
{\fontsize{6}{8}\selectfont
\begin{tabular}{|p{1cm}|p{2.5cm}|p{2.5cm}|p{1.5cm}|p{2.5cm}|p{2.5cm}|p{4.5cm}|} 
\hline
Ref. & Threat Addressed & IDS Datatype & Dataset & Detection Model & XAI Algorithm & Explanation \\
\hline

\cite{nguyen2022enhancing} & DOS, R2L, U2R & Network-based IDS (NIDS) & KDD99, CICIDS2017 & CNN, DT & SHAP & Combines the local and global explanations to improve the interpretation of IDSs. \\
\hline

\cite{roy2022explainable} \cite{das2021machine} & DDoS, XSS and SQL Injection attacks & Network-based IDS (NIDS) & KDD99, CICIDS2017 & Deep Neural Network (DNN), Ensemble models & SHAP, LIME & Generating model-centric and subject-centric explanations from the DNN model's predictions. \\
\hline

\cite{mane2021explaining} & Anomaly detection &  Network-based IDS (NIDS) & NSL-KDD & Deep Neural Network (DNN) & SHAP, BRCG, LIME, ProtoDash, CEM   & Plotting SHAP values to represent low/high values according to their importance. Summarize the trained model by using rules extracted by BRCG. Local explanation for each feature is generated by LIME. Exempler-based explanations by summarizing dataset is generated by ProtoDash. Minimum perturbation possible in the feature value is calculated by CEM. \\
\hline

 \cite{ khan2021new} &  Data injection and Data poisoning attacks in Industrial IoT networks  & Anomaly-based IDS (AIDS)  & Real-world GSP system data source containing time-series dataset  & Conv-LSTM-based autoencoder framework &  LIME &  Illustrate the most relevent attributes and their weights as the basis for interpretation. \\
\hline

\cite{nguyen2019gee} & Low-rate DoS, Port scanning, Botnet, Spam, Blacklist & Cyclostationarity-based NIDS & UGR'16 dataset & Variational Auto-Encoder (VAE) framework & Gradient-based explanation & The interpretability of variational autoencoders is generated by utilizing gradients for clustering anomalies and deriving attack-related fingerprints. \\
\hline
\cite{antwarg2021explaining} & Anomaly detection & Anomaly-based IDS (AIDS) & Warranty claims, KDD Cup 1999, Credit Card Fraud Detection, Artificial dataset & Autoencoder framework & Kernel SHAP & Computes SHAP values for reconstructed features and links them to true anomalous input values to explain prediction errors. \\
\hline
\cite{aguilar2022towards} & Anomaly detection & Anomaly-based IDS (AIDS) & UCI Machine Learning Repository & Decision Tree-based autoencoder & Rule-based Explainability & The correlation values among different categorical attributes provide explanations behind the decision tree. \\
\hline
\cite{lanvin2023towards} & DDoS, XSS and SQL Injection attacks & Network-based IDS (NIDS) & CICIDS2017 & Sec2Graph technique & Explanation based on AE-pvalues & Explanation about the anomaly alert is produced by using the p-value of the empirical distribution of the dimension-wise reconstruction error to flag abnormal feature values. \\
\hline
\cite{javeed2023explainable} & DDoS & Network-based IDS (NIDS) & CICIDDoS2019 & BiLSTM+BiGRU+CNN & SHAP & Using SHAP decision graph including Decision Plot, Waterfall Plot and Summary Plot to demonstrate the important features that contributed most in the detection. \\
\hline
\cite{lin2021towards} & Malware Detection & File Content Analysis & Malimg dataset & Selective Deep Ensemble Learning-based (SDEL) detector & Ensemble Deep Taylor Decomposition (EDTD) & EDTD converting the SDEL prediction to a heatmap, where brighter pixels indicate the more suspicious part in malware binary image. \\
\hline
\cite{IADAROLA2021102198} & Mobile Malware Detection & File Content Analysis & Android Malware Dataset (Argus Lab) & Convolutional Neural Network (CNN) & Grad-CAM & Generating heatmap for visualizing the predictions made by image-based CNN mode. \\
\hline
\cite{andresini2022roulette} & DoS, Probe, R2L, U2R, Fuzzers, Analysis, Backdoors, Exploits, Generic, Reconnaissance, Shellcode, worms & Network-based IDS (NIDS) & NSL-KDD, UNSW-NB15 & Convolutional Neural Network (CNN) & Attention mechanism of ROULETTE & Explainability involves utilizing the attention weights generated by the neural model to provide insights into the classification decisions made by the model for network traffic data. \\
\hline
\end{tabular}
}
\end{sidewaystable}

The explainable and interpretable IDS concept in Industry 5.0 takes on heightened significance as organizations seek to effectively address emerging cyber threats while maintaining transparency and interpretability in their security measures. Explainability in IDSs represents a collaborative effort between AI systems and human operators to address technical challenges at both the model-level implementation and operational levels, enhancing the system's ability to detect and respond to threats. This collaborative approach empowers IDSs to transcend the limitations of black box models by integrating fundamental knowledge and insights, thereby enabling interpretable decision-making processes \cite{bobek2023industry, khan2024need}.

As part of ongoing research and development efforts within the cybersecurity community, traditional intelligent intrusion detection systems are undergoing revisions to incorporate explainability features tailored to diverse stakeholder perspectives \citep{islam2022domain, neupane2022explainable}. These revisions have led to the categorization of explainability into three distinct domains: self-model explainability, pre-modeling explainability, and post-modeling explainability. Self-model explainability encompasses the generation of explanations and predictions in tandem, leveraging problem-specific insights derived from domain expert knowledge. Pre-modeling explainability involves the utilization of refined attribute sets, facilitating clearer interpretations of system behavior before model training. Lastly, post-modeling explainability focuses on shaping the behavior of trained models to enhance their responsiveness to input-output relationships, thereby improving overall system transparency and efficacy in the dynamic landscape of Industry 5.0 cybersecurity \cite{lipton2018mythos}. These types of explanations are discussed in more detail below and their examples are summarized in Table \ref{tab:TABLE II}, \ref{tab:TABLE III}, and \ref{tab:TABLE IV}.\par

\subsubsection{\textbf{Self-model Explainability}}
\label{Self-Model Explainability}
The X-IDS models generated from self-explaining models are designed to inherently explain their intrinsic decision-making procedure. These models exhibit simple architectures capable of identifying crucial attributes that trigger the decision-making process for a given input. In this way, several explainability techniques have been proposed according to the model's complexity, for instance, a Rule-based explanation has been suggested in \cite{sinclair1999application} by developing an Ante-hoc explainability application named NEDAA system that combines ML methods like genetic algorithms and decision trees to aid intrusion detection experts by generating rules, for classifying normal network connections from an anomalous one, based on domain knowledge. The NEDAA approach employs analyst-designed training sets to develop rules for intrusion detection and decision support. In \cite{mahbooba2021explainable}, the authors tried to explain and interpret the known attacks in the form of rules to highlight the target of the attack and their causal reasoning using Decision Trees. They utilized ID3 to construct a decision tree, using the KDD-99 dataset, where the decision rules traverse from top to bottom nodes and the rules from the model are generated using Rattle package from R statistical language. A recent work \citep{manoj2023explainable, 10309785}, proposed decision tree-based explainability to explain the actions taken by the Industrial control system against IoT network activities. These rules can be compiled into an expert system for detecting intrusive events or to simplify training data into concise rule sets for analysts. The rule-based explanation offers valuable insights into decision-making, promotes transparency, and allows domain expertise integration. However, they have limitations in handling complex and evolving threats, scalability, and potential conflicts \cite{neupane2022explainable}. 

Another recent Host-based Intrusion Detection System (HIDS) has been proposed by Yang et al. \cite{yang2023bayesian}, where they used Bayesian Networks (BNs) to create a self-explaining hybrid detection system by combining data-driven training with expert knowledge. BNs are a specific type of Probabilistic Graphical Models (PGMs) that models the probabilistic relationships among variables using Bayes' Rule \cite{pourret2008bayesian}. Firstly, they extract expert-informed interpretable features from two datasets, Tracer FIRE 9 (TF9) and Tracer FIRE 10 (TF10), which consist of normal and suspect system events logs generated through Zeek and Sysmon by the Sandia National Laboratories (SNL) Tracer FIRE team. The authors utilized Bayes Server (2021) as an engine for evaluating multiple BN architectures in finding the best-performing model while the explanations are provided by visualizing the network graph, which provides feature importance information via conditional probability tables \cite{yang2023bayesian}. Self-explaining models in IDS offer notable advantages by enhancing transparency and interpretability. They provide insights into decision-making, enabling analysts to understand the reasoning behind alerts. This aids in trust-building, model validation, and effective response. However, self-explaining models might struggle with complex relationships, limiting their capacity to capture nuanced attack patterns. \par

\subsubsection{\textbf{Pre-modeling Explainability}}
\label{Pre-Model Explainability}
Pre-modeling explainability techniques involve some preprocessing methods to summarize large featured datasets into an information-centric set of attributes that align with human understanding and help downstream modeling and analysis. In \cite{zolanvari2021trust}, the authors proposed explainable model by first transforming the input features into representative variables through Factor Analysis of Mixed Data (FAMD) tool. Then, in the next step, they find mutual information to quantify the amount of informations for each representative and their mutual dependence to the class labels that helps in finding the top explainable representatives for Artificial Neural Network (ANN) model. In the same way, information gain (IG) has been used in \cite{le2022classification} to calculate the most informative feature values, which are then used in Ensemble tree classification. The model's outputs are then plotted in the form of a heatmap and decision plot using the SHAP explanation technique. Also, in \citep{alani2022botstop, ALANI2023103409, Alani2022PAIREDAE, alani2022towards, alani2023explainable, ALANI2023103409} and \cite{alani2023xrecon}, the authors proposed a method named Recursive Feature Elimination (RFE) using feature importance, where the features having lowest importance are removed during the training and testing rounds of different classifiers, including RF, LR, DT, GNB, XGB and SVM classifier. After getting the minimum number of features on which the model shows better performance, i.e., RF, a TreeExplainer which is a type of SHAP explainer, is used to measure the contribution of each selected feature. In a recent work \cite{gurbuz2023explainable}, Gurbuz et al. addressed the security and privacy issue of the IoT-based Healthcare networks data-flow by applying the less computational machine learning models including KNN, DT, RF, NB, SVM, MLP, and ANN. The same procedure of first getting important features using a linear regression classifier and then leveraging the Shapash Monitor explanation interface to visualize feature importance plots, prediction distributions, and partial dependence plots for healthcare professionals, data scientists, and other stakeholders. In \cite{patil2022explainable}, firstly the correlation between features was analyzed using a heatmap, and the outliers were excluded in the preprocessing step. Then LIME technique is used to explain their Voting classifier consisting of RF, DT, and SVM classifiers. In \cite{ zebin2022explainable}, the authors addressed the explainability problem in DNS over HTTPS (DoH) protocol attacks detection system. To understand the underlying distribution of the dataset, the Kernel Density Estimation (KDE) technique has been deployed to estimate the probability density function of the features. After the thorough preprocessing of the datasets, optimal hyperparameters for the base RF classifier has been found by the GridsearchCV function. For the explanation of the model, they used SHAP values to highlight the features that are contributing to the underlying decision of the model. In a recent work \cite{sivamohan2023optimized}, the information-rich features are selected by using the Krill Herd Optimization (KHO) algorithm for BiLSTM-XAI-based classification, where the explanation is provided using both LIME and SHAP mechanisms. In another recent work, Wang et al. \cite{wang2023network} proposed a hybrid explanatory mechanism by first finding the top-most important feature set by using the LIME technique on CNN+LSTM structure. A decision tree model, XGBoost, is then trained on the selected important features while the explanations for the important features are generated through the SHAP mechanism. Another hybrid mechanism has been proposed by Tanuwidjaja et al. \cite{tanuwidjaja2023hybrid} by using both LIME and SHAP mechanisms to cover both the local and global explanations of an SVM-based IDS. 

Another pre-modeling explainability technique involves Visualization, where the focus is on providing intuitive visualizations of data and model behavior to help users, analysts, and stakeholders gain insights into how the model works and why it makes certain predictions. In \cite{mills2019efficient}, a graphical representation has been proposed for understanding the decision tree of the Random Forest (RF) classifier. In the same context, Self-Organizing Maps (SOMs), also known as Kohonen maps \cite{kohonen1982self}, has been used as an exploratory tool to gain a deeper understanding of the data that the decision model is trained on. In \citep{ables2022creating, ables2023explainable}, the authors train and evaluate different extensions of Kohonen Map-based Competitive Learning algorithms including Self Organizing Map (SOM), Growing Self Organizing Map (GSOM) and Growing Hierarchical Self Organizing Map (GHSOM), that are capable of producing explanatory visualizations for the given data. The core designs of these extensions are to organize and represent high-dimensional data in a lower-dimensional space while preserving the topological relationships and structures of the original data. That's why, SOMs can also be used for dimensionality reduction. In terms of IDS explainability, statistical and visual explanations are created by visualizing Global and Local feature significance charts, U-matrix, feature heatmap, and label map through the resulting trained models using NSL-KDD and CIS-IDS-2017 benchmark datasets. Similarly, in \cite{lundberg2022experimental}, the authors proposed a visual explanation method, named VisExp, that applies SHAP to find feature importance values "SHAP-values" for explaining the behavior of the In-Vehicle Intrusion Detection System (IV-IDS). The visual explanation is generated as a dual swarm plot utilizing Python Matplotlib and Seaborn libraries, that presents the normal Controller Area Network (CAN) traffic at the top and the intruder's traffic at the bottom according to the SHAP-values distribution. In \cite{al2022xai}, the authors used the Fast Gradient Sign Method (FGSM) as an adversarial samples generator, and in the next step DALEX framework has been utilized for identifying the most influential features that enhance the Deep Neural Network (DNN) model's decision performance. The same fine-tuning of the deep cyber-threat detection model is also explored by Malik et al. \cite{malik2022xai} by coupling the same adversarial samples generator "FGSM" and the explanations generated through SHAP values. A recent work by Lanfer et al. \cite{lanfer2023leveraging} addresses the false-alarms and dataset limitations issues in available network-based IDS datasets by utilizing the SHAP summary and Gini impurity. Their contribution lies in demonstrating how imbalances in datasets can affect XAI methods like SHAP, and how retraining models on specific attack types can improve classification and align better with domain knowledge. They utilized SHAP Bee-swarm plots to visualize the explanations of the target class individually. 

As such visual explanations offer intuitive insights into the system's behavior but the dependence on visualization quality make the system limited to subjectivity, and may also lead to inconsistency with the changing of visualization technique \cite{zhang2022explainable}. To mitigate these limitations, a combination of various explainability methods, including both visual and non-visual approaches, should be employed to provide a more comprehensive understanding of IDS behaviors and enhance threat detection and prevention capabilities. In \cite{lu2022does}, the authors proposed a feature attribution explainability mechanism based on the concept of an economic theory called Modern Portfolio Theory (MPT). By considering features as assets and using perturbation, the expected feature output attribution values are referred to as their explanation. Feature attribution based on modern portfolio theory minimizes the variance of prediction score changes about attribution values, indicating that a higher feature attribution value signifies a substantial impact on the model's prediction score with a small feature change. \par

\subsubsection{\textbf{Post-model Explainability}}
\label{Post-Model Explainability}
Post-model explainability refers to the techniques and methods used to interpret and understand the decisions made by a trained learning model. Unlike self-and pre-model explainability techniques, post-model allows stakeholders to gain insights into model decisions, detect biases, and validate model behavior, contributing to better-informed decision-making and building trust in AI systems \cite{9872110}. The most adopted techniques in the literature include the feature importance methods, where the impact of each input feature is analyzed according to the trained model's performance. In \cite{SARHAN2022100359}, the authors used the SHAP method to explain the ML model detection performance. After finding the best-performing sets of hyper-parameters for both the MLP and RF classifiers through partial grid search, the models are analyzed to understand their internal operations by calculating the Shapley value of the features. The LIME explainability approach for detecting adversarial attacks on IDS systems has been proposed in \cite{tcydenova2021detection}, where the normal data boundaries are explained for a trained SVM-based model. In the same way, in \cite{gaitan2023explainable}, horizontal bar plots are used to visualize the global explanation of the model prediction using the SHAP mechanism. Another work by Oseni et al. \cite{oseni2022explainable}, also proposed a SHAP mechanism for improving interpretation and resiliency of the DL-based IDS in IoT networks. The SHAP mechanism has also been proposed by Alani et al. \cite{alani2022deepiiot,alani2023arp} and Kalutharage et al. \cite{kalutharage2023explainable}, to explain a Deep-learning-based Industrial IoT (DeepIIoT) intrusion detection system. 

Along with employing LIME and SHAP mechanisms to explain the prediction made by the Extreme Gradient Boosting (XG-Boost) classifier, the authors in \cite{muna2023demystifying} also used ELI5, "Explain Like I'm 5", a python package using the interpreting Random Forest feature weights approach. This package supports tree-based explanation to show how effective each feature is contributing on all parts of the tree in the final prediction. Abou et al. \cite{abou2022novel} used RuleFit and SHAP mechanisms to explore the local and global interpretations for DL-based IDS models. In \cite{ Marino2018AnAA}, an adversarial ML approach is utilized to find explanation for an input features. They used the samples that are incorrectly predicted by the trained model and tried again with the required minimum modifications in feature values to correctly classify. This allowed the generation of a satisfactory explanation for the relevant features that contributed to the misclassification of the MLP model. The same idea has been deployed in \cite{da2023false}, where the authors combined the SHAP and Adversarial approach to accurately identify the false positive prediction by the IDS model. In \cite{szczepanski2020achieving}, the authors proposed a prototype system where they utilized Feed Forward ANN with PCA to train as a classifier and in parallel a decision tree is generated from the samples along with their outputs from the classifier. The retrieved tree is handled by the DtreeViz library to visualize an explanation for the classifier's decision. In \cite{wang2020explainable}, the authors improve the explanation of an IDS by combining local and global interpretation generated by models using the SHAP technique. Local explanation gives the reason for the decision taken by the model, and the global explanation shows the relationships between the features and different attacks. They used the NSL-KDD dataset and two different classifiers, namely, one-vs-all and multiclass classifiers are utilized to compare the interpretation results. The same mechanism has been adopted in \cite{nguyen2022enhancing}, to explain the decision made by CNN and DT-based IDS model using the SHAP values. The target is to build trust for the design of the intrusion detection model and security experts. In \cite{roy2022explainable} \cite{das2021machine}, a SHAP-LIME hybrid explainability technique has been proposed to explain the generated results by the DNN both globally and locally. The same hybrid approach is also presented in \cite{mane2021explaining} for explaining a deep neural network-based IDS. To provide quantifiable insights into which features impact the prediction of a cyber-attack and to what extent, they used the SHAP, LIME, Contrastive Explanation Method (CEM). ProtoDash and Boolean Decision Rules via Column Generation (BRCG) approaches.

Learning the compact representations of input data through the encoding and decoding process, Autoencoders aid in uncovering underlying patterns and essential features within the data. This latent representation often corresponds to meaningful characteristics of the input data, making it easier to understand and interpret the model's behavior \cite{charte2020analysis}. The autoencoders are based on reconstructing the input samples by minimizing the reconstruction error between the encoder and decoder. Along with the great property of anomaly detection, the reconstruction-error-based methods also provide a comprehensive explanation of the connection between the inputs and the corresponding outputs. In this context, Khan et al. \cite{ khan2021new} proposed an autoencoder-based IDS architecture by adopting CNN and LSTM-based autoencoder to discover threats in the Industrial Internet of Things (IIoT) as well as to explain the model internals. For the model explainability, the LIME technique has been used to explain the predictions of the proposed autoencoder-based IDS. The same LSTM-based autoencoder model for anomaly detection in industrial control systems has been proposed in \cite{HA20221183}, where the explainability of the model is achieved by the Gradient SHAP mechanism. Another work \cite{nguyen2019gee}, used a variational autoencoder (VAE) to detect network anomalies and a gradient-based explainability technique to explain the models' decisions. In \cite{antwarg2021explaining}, the authors used the reconstruction error as an anomaly score and computed the explanation for prediction error by relating the SHAP values of the reconstructed features to the true anomalous input values. In the same way, \cite{aguilar2022towards} proposed a decision-tree-based interpretable autoencoder, where the correlation between the categorical attribute's tuples are learned through the decision-tree encoding and decoding process and the interpretability of the autoencoder is seen by finding the rules from the decoder to interpret how they decode the tuple accurately. A recent research work \cite{lanvin2023towards}, proposed a novel explainability mechanism named AE-values, where the explanation is based on the p-values of the reconstruction errors produced by an unsupervised Auto-Encoder-based anomaly detection method. They handle the anomaly detection problem as a one-class classification problem using the Sec2graph method and a threshold value is computed from the reconstruction error of the input benign files. The error value above the threshold is considered responsible for the anomaly. Another recent work, \cite{javeed2023explainable} proposed a multi-class prediction model by combining BiLSTM, Bidirectional-gated recurrent unit (Bi-GRU), and fully connected layers. They applied the SHAP mechanism on the last fully connected layer to get the local and global interpretation for the model decision.  \par

\begin{figure*}[ht]
\centering
\includegraphics[width=\linewidth]{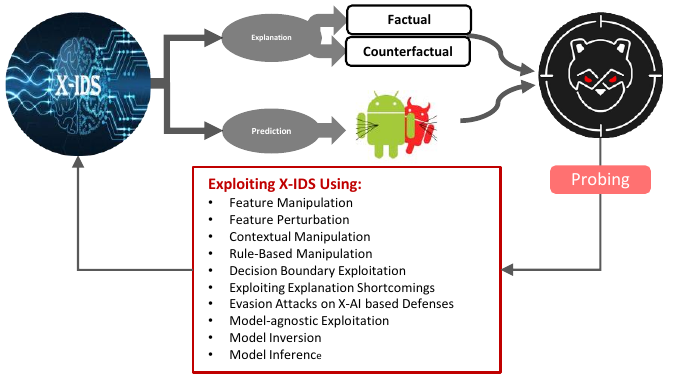}
\caption{In X-IDS (Explainable Intrusion Detection Systems), the specific exploitation and manipulation techniques can vary based on the X-IDS implementation, and the effectiveness of these techniques depends on the quality of explanations provided and the adversary's knowledge of the system.}
\label{fig: Adv-XAI}
\end{figure*}

Another post-model explainability technique involves Saliency Map or Attention Map methods, which aim to explain the decisions of a Convolutional Neural Network (CNN) by highlighting the regions of an input image that contribute most to a specific prediction. In \cite{yoon2015memory}, a method named Memory Heat Map (MHM) has been proposed to characterize and segregate the anomalous and benign behavior of the operating system. In \cite{lin2021towards}, the authors used the region perturbation technique to generate a heatmap for visualizing the predictions made by the image-based CNN model. Idarola et al. \cite{IADAROLA2021102198} proposed a cumulative heatmap generated using the Gradient-weighted Class Activation Mapping (Grad-CAM) technique, where the gradients of the convolutional layer are converted into heatmap through the Grad-CAM to balance the trade-off between CNN accuracy and transparency. In \cite{andresini2022roulette}, the network traffic classification and decision explanation task are addressed through image classification, where the network flow traces are transformed into a pixel frame of single-channel square images. A CNN model is incorporated with the attention layer to capture the contextual relationships between different features of the network traffic and the observed intrusion classes.  These techniques help users understand not only which regions are important but also the extent to which different regions influence the model's predictions. It can be particularly useful for gaining a finer-grained understanding of the relationships between input features and model responses \cite{hariharan2021explainable}. \par
From our extensive review of the literature, it becomes evident that across various Machine Learning (ML)-based Intrusion Detection Systems (IDS), efforts have been directed towards attaining both local and global explainability of the model's decision. This has been achieved through the utilization of Rule-based approaches, Local Interpretable Model-agnostic Explanations (LIME), and SHapley Additive exPlanations (SHAP) techniques. However, it is noteworthy that while these techniques, LIME and SHAP, have shown effectiveness in other domains, such as image data interpretation, the nature of IDS problems presents unique challenges. In IDS, attributes are often highly interrelated and interdependent, requiring an approach that can accurately capture these complex relationships in its explanations. Therefore, there exists a pressing need for the development of more efficient techniques tailored to the specific complexities of IDS, ensuring that explanations are not only interpretable but also reflect the nuanced interactions within the data.


\section{\textbf{Adversarial XAI and IDS}}
\label{sec:adversarialIDS}
The adaptability of DL-based systems in cybersecurity within the context of Industry 5.0 is not yet mature compared to other domains, such as instance recognition systems, recommender systems for business and social platforms, etc. There are two main reasons. The first one is the lack of confidence due to the black-box nature of these intelligent models \cite{charmet2022explainable}. To aim, efforts are made for explainable AI systems, where a model can reason their decision by providing a reasonable explanation. This resulted in the recent trend of white-box AI models by generating model-specific and model-agnostic explanation techniques \cite{duddu2018survey} \cite{dwivedi2023explainable}. The second reason, which is the focus of this section, is the \textit{Adversarial methodologies}, where an intelligent model is made fool purposely by generating adversarial examples that the model can not predict correctly \cite{ling2023adversarial}. These kinds of attacks exist for both the black-box and white-box models, where the factual and counterfactual explainability generated by the XAI methods make this adaptability more insecure and provide an easy way to launch various adversarial attacks on the models, shown in Figure \ref{fig: Adv-XAI}, such as feature-manipulation/perturbation, Decision Rules-manipulation, evasion, membership inference, model poisoning, or vulnerability extraction attacks \cite{shokri2017membership}. \par

\begin{table*}[htbp]
\scriptsize
\centering
\caption{\textbf{Adversarial XAI techniques}}
\label{tab:TABLE V}
{\fontsize{6}{8}\selectfont
\begin{tabular}{|p{0.5cm}|p{1.8cm}|p{2.5cm}|p{1.6cm}|p{2.0cm}|p{2.8cm}|}
\hline
\textbf{Ref.} & \textbf{Data type} & \textbf{Dataset} & \textbf{Attack Type} & \textbf{Detection Model} & \textbf{XAI targeted/Non} \\
\hline
\cite{piplai2020nattack} & Network events logs & IEEE BigData 2019 Cup: Suspicious Network Event Recognition & Perturbation & GAN & \ding{55} \\
\hline
\cite{ayub2020model} & Network-based & CIC-IDS-2017, TRAbID 2017 & Perturbation & MLP & \ding{55} \\
\hline
\cite{alshahrani2022adversarial} & Network-based & CIC-IDS-2017 & Evasion & DT, LR & \ding{55} \\
\hline
\cite{duy2023investigating} & Network-based & CIC-IDS-2018, InSDN & Evasion & DT, LR, CNN, MLP, LSTM & \ding{55} \\
\hline
\cite{zhang2020brute} & Host-based, Network-based, Application-based & ADFA-LD, NSL-KDD, DREBIN & Perturbation & DT, LR, MLP, NB, RF & \ding{55} \\
\hline
\cite{qiu2020adversarial} & IoT Network-based & Mirai, Falsifying Video streaming application & Perturbation & DNN-based Auto-encoder & \ding{55} \\
\hline
\cite{chen2021fooling} & Network-based & NSL-KDD, UNSW-NB15, CICIDS2017 & Evasion & Auto-encoder & \ding{55} \\
\hline
\cite{jiang2022fgmd} & IoT Network-based & MedBIoT, IoTID & Perturbation & LSTM, RNN & \ding{55} \\
\hline
\cite{ravikrishnan2023ardl} & Network-based & KDDCup'99 & Evasion & DNN & \ding{55} \\
\hline
\cite{li2023eifdaa} & IoT Network-based & X-IIoTID & Evasion & SVM, DT, RF, KNN, CNN, GRU, HyDL-IDS & \ding{55} \\
\hline
\cite{debicha2023adv} & Network-based & CTU-13, CSE-CIC-IS2018D & Evasion & MLP, RF, KNN & \ding{55} \\
\hline
\cite{li2018bebp} & Host-based & KDDCUP99, NSL-KDD, Kyoto 2006+ & Poisoning & NB-Gaussian, LR, SVM-sigmoid & \ding{55} \\
\hline
\cite{xu2020approach} & Network-based & LANL network security dataset & Poisoning & LSTM, B-LSTM, T-LSTM & \ding{55} \\
\hline
\cite{nguyen2020poisoning} & Network-based & D¨IoT-Benign, UNSW-Benign, D¨IoT-Attack & Poisoning & Federated Learning-based DNN & \ding{55} \\
\hline
\cite{rosenberg2020generating} & Portable Executable (PE) & EMBER & Adversarial examples & GBDT & Integrated Gradients, DeepLIFT, LRP \\
\hline
\cite{zhang2024explainable} & Network-based IDS & CIC-IDS2017, Kitsune & Adversarial examples & MLP, AlertNet, IDSNet, DeepNet, RF, Xgboost, MaMPF, FS-Net, KitNET, Diff-RF & SAGE (SHAP) \\
\hline
\cite{kuppa2020black} & Portable Executable (PE), Network-based & Malicious/Benign PDF files, Android Apps, UGR16 & Perturbation & MLP, Adversarial Auto-encoder & gradient-based XAI \\
\hline
\cite{kuppa2021adversarial} & Network-based, Portable Executable (PE) & Leaked Password, CICIDS17, VirusShare & Evasion, Membership inference, Poisoning, Model extraction & Auto-encoder, GBM, NN & Latent Counterfactual, Permute Attack, Diverse Counterfactual \\
\hline
\cite{severi2021explanation} & Portable Executable (PE) & EMBER, Contagio (PDFs), Drebin (Android executables) & Evasion, Membership inference, Poisoning, Model extraction & Auto-encoder, GBM, NN & Latent Counterfactual, Permute Attack, Diverse Counterfactual \\
\hline
\cite{ruijin2022instance} & Portable Executable (PE) & Microsoft Malware classification Challenge & Evasion & DNN & Superpixels \\
\hline
\cite{alani2023adversarial} & Network-based & Iot network intrusion dataset & Evasion & XGB & SHAP \\
\hline
\cite{shu2023eagle} & Portable Executable (PE) & Drebin (Android executables) & Evasion & RF, MLP & LIME \\
\hline
\end{tabular}
}
\end{table*}

Explainability in cybersecurity plays a double-edged sword and very little work has been done to enhance the robustness of explainable models. From the above X-IDS section \ref{sec:Intrusion Detection Systems for Cybersecurity in Industry 5.0}, and also from Table \ref{tab:TABLE II}, \ref{tab:TABLE III}, \ref{tab:TABLE IV}, it can be seen that the prominent competing methods in explainability are the explanation of the model through the coefficients of regression models, Rule-based, LIME, SHAP or Gradient-based explanations, etc. The majority of these techniques are evaluated based on descriptive accuracy and relevancy characteristics of the XAI models \cite{hall2019systematic}. The objective characteristic sets are divided based on XAI design goals and evaluation measures among target users, such as XAI models for AI novices, Data analysts, and AI design experts and the evaluation measures include useful and satisfactory explanations, versatile and effective information collection and computation cost explainability \cite{mohseni2021multidisciplinary}. In the context of cybersecurity, having enough information about the internal decision-making system, an attacker can deceive both the target security model and their explainability method. This vulnerability underscores the critical need for robust defenses against adversarial attacks in the development and deployment of AI-based cybersecurity systems in Industry 5.0. Therefore, there is a pressing need for further research to explore robust and adversarial-resistant explainability methods. These methods are essential to safeguard XAI-based anomaly detection and prevention systems against malicious attacks \cite{srivastava2022xai}. \par

Before the advent of ML-based solutions, network anomalies that might imply an attack were identified through well-crafted designed rules. In situations where attackers possessed expertise in cybersecurity, they could intelligently infer which specific features of network traffic data were under scrutiny by the cyber-defense mechanism. Armed with this knowledge, the attacker could easily circumvent a rule-based cyber-defense system \cite{chakraborty2021survey}. However, the intelligent ML-based solutions for network intrusion detection systems show promising results in mitigating these threats to some extent, but the everlasting competition between attackers and defenders results in the innovation of muddling strategies to defeat each counter mechanism. Models composing deep processing layers to learn high-level abstraction from past experiences and to perceive the future in terms of conceptual hierarchy can also be victimized adversarially and could lead to forcing the model's behavior contrary to their proposed intended functionality \cite{sarker2023multi}. The literature reveals numerous compelling studies focused on adversarial attacks against IDS. In the subsequent sections, we delve into adversarial attacks and their implications for IDS. To enhance clarity, our discussion is structured into two subsections: adversarial attacks conducted without and with the utilization of explainability mechanisms.

\subsection{\textbf{Adversarial Attacks without utilizing Explainability}}
Adversarial attacks on ML can be categorized into three types: white-box attacks, where the attacker has complete knowledge of the target system; black-box attacks, where the attacker lacks knowledge about the detection mechanism but can query the model to gain information; and gray-box attacks, where the attacker has limited information about the classifier, such as some features or the algorithm used without configuration details \cite{apruzzese2019addressing}. In terms of offensive use of inference models, the possible attack vectors include privacy attacks (e.g., membership inference, model inversion, and model extraction), poisoning attacks (e.g., backdoor injection), and evasion attacks (e.g., test-time adversarial perturbations). The most easy and effective way of attacks involve perturbation and evasion mechanisms in the context of machine learning and artificial intelligence. Perturbation attacks aim to subtly alter input data to mislead machine learning models, while evasion attacks aim to create inputs that completely bypass model detection \cite{alatwi2021adversarial} \cite{apruzzese2022modeling} \cite{merzouk2021deeper}. 

To mitigate the adversarial attacks, research studies also proposed methodologies for training the model on adversarial examples to counter malicious activities before being attacked. However, these solutions also proved a failed encounter. The most significant methodology involves Generative Adversarial Networks (GAN) and their respective variants, which have gained substantial traction across various domains of ML applications, including cybersecurity, due to their ability to generate synthetic data, address the class imbalance, create adversarial examples, and enhance the manipulation of data's semantic information \cite{dunmore2023comprehensive} \cite{alkadi2023better}. In \cite{piplai2020nattack}, the authors target attack a proposed solution based on GANs for countering the adversarial attacks on AI-based models. They used the discriminator neural network part of GAN as a trained classifier which can effectively distinguish between fake and real inputs and then they successfully attacked the trained classifier by perturbing malicious samples using the Fast Gradient Sign Method (FGSM), which is a method to calculate the gradient of the loss function. The same perturbation attack mechanism has been used by Ayub et al. \cite{ayub2020model} by utilizing the saliency map to iteratively modify input dimensions for a targeted MLP model to create adversarial samples. They used the Jacobian-based Saliency Map Attack (JSMA) method to generate adversarial test samples with slight perturbations in legitimate data. Pujari et al. \cite{pujari2022approach} used the Cerlini-Wagner (CW), a powerful white-box evasion attack technique, to generate adversarial samples for evaluating their trained GAN-based IDS mechanism. In \cite{alshahrani2022adversarial}, the authors targeted two most commonly used classifiers in IDS taxonomy (i.e., DT and LR) to perform evasion attack by generating synthetic samples using Deep Convolutional Generative Adversarial Network (DCGAN). Besides the effectiveness of GANs in generating latent patterns in the case of detecting zero-day attacks, the instability in training results in introducing other GAN variants. In \cite{duy2023investigating}, the authors investigated the weaknesses of ML-based IDS in software-defined networks (SDN) environments by generating adversarial samples using three GAN variants including Wassertein GAN with gradient penalty (WGAN-GP), WGAN-GP with two timescale update rule (WGAN-GP TTUR), and AdvGAN. They targeted to perturb specific non-functional features in the SDN environment to evade detection of malicious traffic flow. In \cite{zhang2020brute}, the authors evaluated the robustness of an ML classifier by a gradient-free method. Instead of trying the GAN-based attack, they tried to manipulate input features based on confidence scores from target classifiers to create adversarial examples, using the scores to guide feature modifications. 

Similarly, membership inference attacks pose a severe threat to the privacy and security of AI models by potentially revealing information about the training data, compromising confidentiality. In \cite{shokri2017membership}, the authors introduced a membership inference attack by developing their surrogate predictive models, one for each target model's output class as a shadow model, trained on synthetic data generated based on confidence value shown by the target model against the data sample. This shadow training technique leverages the similarity of models trained on similar data records, allowing for effective inference of membership in the target model's training dataset, based on knowledge of the shadow model's training dataset. In the same way, Qiu et al. \cite{qiu2020adversarial} presented a novel black-box adversarial attack on DL-based Network Intrusion Detection Systems (NIDSs) in IoT environments. The authors employed model extraction and saliency maps to reveal critical packet attributes and efficiently generate adversarial examples, achieving a high attack success rate of 94.31\% while modifying less than 0.005\% of the bytes in malicious packets. In \cite{chen2021fooling}, the authors introduced the Anti-Intrusion Detection Auto-Encoder (AIDAE), a novel feature generative model designed for learning the distribution of normal features and generating random features that can mimic real network traffic flows. The AIDAE employs multi-channel decoders to generate both the continuous and discrete features, ensuring that the generated features maintain the correlation between these components and closely match the distribution of normal features. This approach aids attackers in bypassing existing Intrusion Detection Systems (IDSs). To counter adversarial attacks against a black-box model, Jiang et al. \cite{jiang2022fgmd} first proved a successful perturbation attack against the two deep learning-based IDS models, LSTM and RNN, and then proposed the defensive mechanism, Feature Grouping and Multi-model fusion Detector (FGMD), to improve the robustness against the targeted adversarial attacks. Ravikrishna et al. \cite{ravikrishnan2023ardl} deployed different gradient-based evading attack techniques including Fast Gradient Sign Method (FGSM), Basic Iterative Method (BIM), Momentum Iterative Method (MIM) and Projected Gradient Decent (PGD) to expose the loopholes in the detection model and then subsequently retrained the DNN-IDS with the shuffled batch of adversarial and normal samples to enhance the performance against evading attacks. In \cite{merzouk2022evading}, the authors target a Deep Reinforcement learning-based NIDS with adversarial examples generated by FGSM and BIM to evade detection. The same mechanism is adopted in \cite{li2023eifdaa}, where FGSM, BIM, PGD, Deepfool, and WGAN-GP adversarial attack mechanisms are used to generate AEs and retrain the ML-based IDS to improve their robustness.   

Within the domain of black-box attacks, two key methodologies have emerged. These methodologies include model querying, involving the issuance of queries to the target model for insights to extract useful information allowing crafting perturbation attack, and transferability, which explores the potential for adversarial samples crafted for one model to be effective with a high probability against another. Transferability attacks present a significant threat to AI models, allowing adversaries to exploit vulnerabilities in one model and leverage the acquired knowledge to compromise other related models, undermining their security. Debicha et al. \cite{debicha2023adv} adopted the transferability technique where they target NIDS by first generating adversarial traffic specifically designed for the surrogate models to bypass. Then, they target the defender NIDS to attack by sending those adversarial botnet traffic that managed to avoid detection by the surrogate models. The continuous data collection and training requirements of DL-based abnormal threat detection systems make them susceptible to poisoning attacks. Poisoning training data with the intention of causing misclassification of specific instances, degrading model accuracy, or introducing vulnerabilities that can be exploited later during testing, is also one of the crucial mechanisms in IDSs. Li et al. \cite{li2018bebp} introduced a novel poisoning method, leveraging the Edge Pattern Detection (EPD) algorithm, to exploit different machine learning models. They employed a boundary pattern detection algorithm to create poisoning data points that closely resemble normal data but evade current classifiers. Furthermore, they enhance the technique with a Batch-EPD Boundary Pattern (BEBP) detection algorithm to yield more valuable boundary pattern points, leading to a moderately effective poisoning method known as the chronic poisoning attack. Xu et al. \cite{xu2020approach} targeted LSTM using a poisoning attack. They adopted a discrete adversarial sample generation method to create a replaceable set for each feature and generate samples through feature substitution from the set rather than introducing disturbances in training data. Similarly, Nguyen et al. \cite{nguyen2020poisoning} targeted a federated learning-based IDS system by getting access to a compromised security gateway and injecting targeted malicious traffic patterns as well as benign traffic to corrupt the global model. 

\begin{figure*}[ht]
  \centering
  \begin{tikzpicture}
    \node[inner sep=0] (image) at (0,0) {\includegraphics[width=1\textwidth, height=.3\textheight]{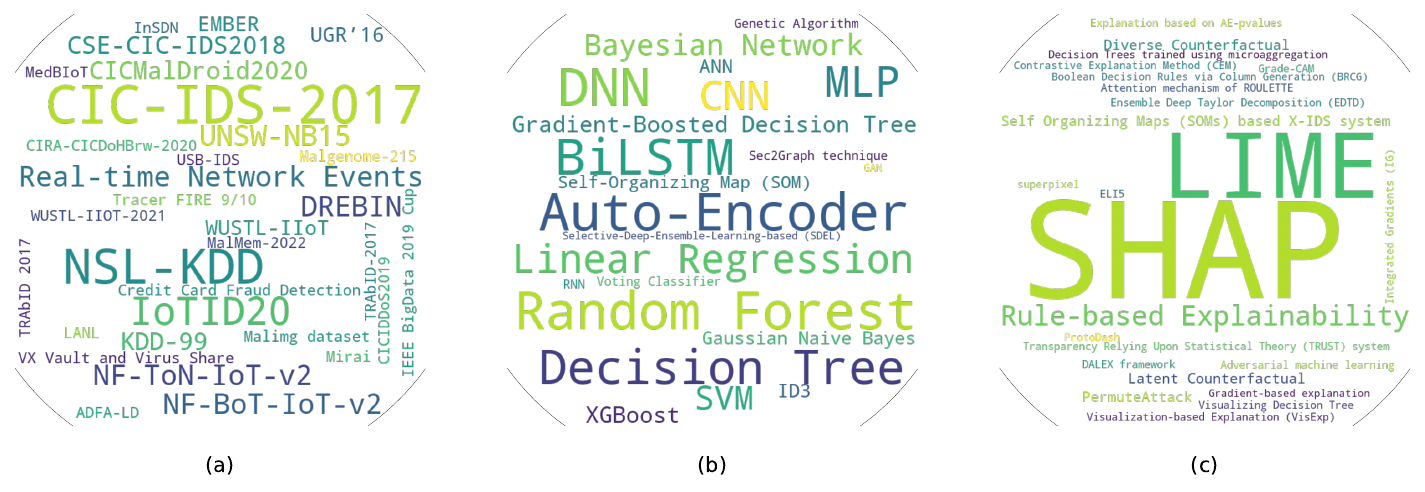}};
    \draw[line width=0.5pt, color=black!20] (image.south west) rectangle (image.north east);
  \end{tikzpicture}
  \caption{Summarize the three main columns of Table \ref{tab:TABLE II}, \ref{tab:TABLE III}, \ref{tab:TABLE IV} and \ref{tab:TABLE V}. In Figure A, we make the word cloud of Datasets used mostly in the literature, Figure B shows the most used Detection models for IDS problems, and Figure C shows the most deployed Explainability techniques specifically used for IDSs. }
  \label{fig:wordcloud}
\end{figure*}

Most of the above-discussed adversarial attacks involve manipulating input data to deceive the black-box model, potentially evading detection or forcing it to generate false decisions. Besides the successful attack scenarios shown in the literature, there are limitations in the case of real environment \cite{he2023adversarial}. As we know IDSs are typically label-only black-box systems, and the adversary generally does not know the feature dimension nor the gradients and confidence scores of the target model. It only outputs a binary decision of whether the input is benign or malicious. These limitations makes most of the attack scenarios impractical and need more technical strategies to bypass the decision model. Another key aspect is functionality preservation such that any crafted attack can still execute the original intended functionality when inspected by a human and/or a machine \cite{mccarthy2022functionality}. The exposure to attack becomes easier when the attacker gains insights into the black-box operations using explainability techniques. Along with the benefits of transparency, enhanced human understanding of alerts, and the ability to uncover novel attack patterns, the risk of revealing the system's architecture and vulnerabilities to attackers make the decision model more prone to adversarial attacks \cite{tcydenova2021detection}. \par

\subsection{\textbf{Adversarial Attacks with utilizing Explainability}}
Some of the XAI methods have been deployed to comprehend, identify, and defend against specific adversarial attacks. Most of the employed approaches involve developing visualizations to highlight the regions vulnerable to changes or could most likely be altered by adversaries. In \cite{tcydenova2021detection,rehman2023explainable}, the LIME technique is employed to anticipate the attacks and normal traffic data. The method involves finding the most important feature sets for normal traffic and using those features as a comparison set for the detected normal traffic to validate the model's decision. Though these approaches work well in identifying potential vulnerabilities in the model, the explanation methods are still incapable of detecting various adversarial attacks and can be manipulated to affect the user's trust or be exploited to launch different attacks. 

The trade-off between X-IDSs and adversarial attacks is complex and multifaceted, which is why some most popular explainability mechanisms are used as counterintuitive. Exploiting XAI's enhanced capabilities in the lens of classical CIA triad, \textit{Confidentiality} attacks employ explanations to reveal model architecture or training data while \textit{Integrity} and \textit{Availability} attacks use explanations to uncover insights enabling adversaries to manipulate model results or disrupt access to legitimate users \cite{nadeem2023sok}. These attacks can occur during training (e.g., poisoning) or deployment (e.g., evasion), depending on the attacker's strategy, timing, and objectives. \par

The advantage of explainability in highlighting the critical decision boundaries could also be exploited by adversaries. For instance, In \cite{rosenberg2020generating}, the authors introduced the concept of transferability of explainability, a similar concept of adversarial example transferability, where the impactful features are identified utilizing explainability algorithms on a substitute model with the assumptions that their impact will be same on any target black-box model. Specifically, they used Kendall's tau, a statistical measure (often used for comparing rankings), to compare the rankings of features produced by different explainability mechanisms. Being unaware of the attacked classifier, this transferability of explainability can help the adversary in generating adversarial examples by modifying some structural features without affecting their core functionality. A recent work \cite{zhang2024explainable} targets the same concept of transferability by proposing a novel method named Explainable Transfer-based black-box adversarial Attack (ETA) framework. The ETA framework optimizes a substitute model, selects important sensitive features, and crafts adversarial examples guided by gradient evaluation, enhancing transferability against target NIDSs.

In \cite{kuppa2020black}, the authors proved the possibility of compromising the confidentiality of the target classifiers and their explainability mechanisms. The work employs the Manifold Approximation Algorithm (MAA) to discern patterns in data distributions, subsequently generating synthetic data boundaries. These boundaries are then perturbed to probe an ML classifier, and explanations are obtained to understand its behavior in a black-box setting. Kuppa et al. \cite{kuppa2021adversarial} presented four successful black-box attacks including evasion, membership inference, poisoning, and model extraction attacks by utilizing three different counterfactual explanation methods to compromise the confidentiality and privacy of the target classifiers. They targeted generating counterfactual explanation-based synthetic malware samples by modifying static features, for example, adding a few debugging information bytes or changing section names, without affecting their functionality. Another insidious backdoor poisoning attack on feature-based malware classifiers has been proposed in \cite{severi2021explanation}, where the SHAP explainability mechanism is used to select a highly effective set of features and their values as a backdoor for the victim model during training time. This approach aims at generating modified benign samples with embedded backdoors, distributing them through labeling platforms to contaminate the malware classifier's training dataset, and subsequently inserting the same watermark into malicious binaries to ensure evasion of detection by the backdoored classifiers during inference. In \cite{ruijin2022instance}, the authors proposed a binary-diversification technique, named FastLSM (Fast Least Square Method), to transform the influential data sections of the malware file to make them resemble benign classes, thereby achieving evasion. They leverage Superpixels, an image segmentation technique, as an interpretation mechanism for a disassembled malware binary file's basic functional blocks, which also aids in perturbing specific code subsections of the file to fool DNN models. A recent work \cite{alani2023adversarial}, proposed an evasion attack for a Botnet black box detection system utilizing the SHAP explainability mechanism to generate adversarial samples. The most impactful SHAP values are targeted for each feature having maximum contribution in the benign decision. These features are then used for successfully generating adversarial samples. In \cite{shu2023eagle}, the authors target the count features in the sample file to evade different count feature-based Android malware classifiers. They used the LIME explainability mechanism for the important features positively contributing to the classification process. New variants are then generated by modifying those important feature values to alter the model's classification decisions.

\section{\textbf{XAI-based IDS: Lessons Learned, Challenges and Future Research Directions}}
\label{X-IDS Challenges and Future Research Directions}
The emerging concept of Industry 5.0 has received great attention from the research community. It includes Human-Machine Interaction, Cyber-physical systems, Robotics and Automation, Industrial Internet of Things (IIoT), and Big Data Analytics using AI and ML. Within this spectrum of concepts, the privacy and security of the information exchange systems assume a vital role, influencing the establishment of trust among diverse stakeholders and facilitating the adoption of these technological transitions. 

In this scientific review, we tried to cover the recent advancement in cybersecurity specifically targeting a comprehensive survey on IDSs with the advancement of explainable IDS. Current X-IDSs have significantly improved the interpretability and transparency of AI-based intrusion detection and prevention systems. These mechanisms aim at uncovering the black-box nature of the AI-based models and showed fruitful progress from different stakeholders' perspectives as shown in Table \ref{tab:TABLE II}, \ref{tab:TABLE III}, \ref{tab:TABLE IV}. However, they still face some critical limitations including complexity, scalability, the trade-off between explainability and accuracy, lack of standardization, and most importantly the adversarial attacks as shown in Table \ref{tab:TABLE V}. 

According to the literature study, the generic explainability mechanisms include two main categories, ante hoc and post-hoc explainability. The ante hoc explainability mechanisms include models that generate explanations and predictions together, which are also called Self-explaining models. Due to simple architecture with problem-specific notions, the decision of these models is their explanation. For example, Rule-based explanations are generated by the domain experts by designing the decision tree of the actual system. Similarly, post-hoc explainability mainly captures the relationship between the input instances and the output of a complex black-box model. This type of explanation methods are further categorized into two classes, namely, pre-modeling and post-modeling explainability mechanisms. These methods are either model-specific or model-agnostic depending on the specific use case. 

After carefully investigating the recently proposed explainability mechanisms, it is clear that the current industry 5.0 revolution and research community strive to unveil the complex patterns in the deployed AI-based cybersecurity decision models. This could improve the trade-off between model explainability and accuracy. This prevalence results in the development of XAI-based IDS systems in the cybersecurity field. As we can see in Figure \ref{fig:wordcloud}, a diversified set of sources including datasets, ML models, and explainability mechanisms are used by the research community to address this problem. Along with investigating these various sources and mechanisms, we also tried to cover the adversarial X-IDS mechanisms, where attackers exploit the available explanation to attack the X-IDS model. 

\subsection{Challenges in Developing XAI-Based IDS Systems and the Way Forward}
The development of Explainable Intrusion Detection Systems (X-IDSs) poses multiple challenges mainly stemming from the inherent complexity of modern information exchange protocols, network architectures, and the sophisticated nature of cyber attacks in the Industry 5.0 realm. One significant challenge is coping with the complexity of deep learning models employed in X-IDS. While these models demonstrate exceptional performance, their opacity poses a hurdle in terms of interpretation. Extracting meaningful explanations from these complex models demands sophisticated techniques, often requiring a trade-off between model accuracy and interpretability. 

Due to the simplicity and incapability of ante-hoc explainability mechanisms to effectively capture the morphed patterns of modern cyber threats, the current literature is focused on finding the relevancy between the input sample features and their output by adopting different Post-hoc mechanisms. The majority of these methods are mostly rooted in image processing, and their applicability in AI-based cybersecurity problems doesn't fit well because of the complex nature of the data, which has sequential patterns, and textual and categorical features. These simple feature attribution and saliency mapping mechanisms can not accurately capture the complex relationship within the data. Moreover, the integration of contextual information, such as user behavior and network context, adds another layer of complexity. Despite the significant efforts in the domain, the question about "\textit{\textbf{How an explainability mechanism can be defined in IDS systems which can effectively interpret the temporal and contextual dependencies for a specific cyber threat?}}" remains unsolved. X-IDS must confront the dynamic and ever-evolving nature of cyber threats, necessitating the incorporation of contextual understanding to enhance decision-making.

Moreover, addressing adversarial attacks is imperative, as threat actors continuously seek to exploit vulnerabilities in IDS systems. Ensuring robustness against adversarial manipulations and generating explanations capable of discerning authentic threats from adversarial instances poses a formidable challenge. Explainability in the cybersecurity domain plays a double-edged sword. For example, if adversaries gain insights into the decision model, they can deceive both the target security model and the explainability method. The adversarial explainable intrusion detection systems (Adv-X-IDS) pose a critical challenge for security analysts. How can we effectively integrate explainability and interpretability into an AI-based black-box model, particularly when considering adversarial scenarios, to enhance the generalization of the decision process within the framework of the CIA triad in cybersecurity? needs to be yet addressed. The intricacies of adversarial attacks in the context of XIDS demand innovative solutions to enhance the resilience of intrusion detection models and fortify the interpretability mechanisms against intentional subversion by malicious actors.

The scalability of X-IDS systems is also a concern, especially when dealing with large-scale network data in real time. Balancing efficiency without compromising the depth of explanations becomes crucial in the face of evolving cyber threats. Lastly, ethical considerations, user-friendly explanations, and effective integration with human expert decision-makers further increase the complexity involved in the design of X-IDS systems.

\section{Conclusion}
\label{sec:conclusions}
The obscure nature of the complex AI methods raises the need for in-depth evaluation of the decisions made by the Deep Learning models. X-AI has recently introduced the concept of White-box models, allowing the interpretation of internal information and decisions in AI-based systems. Similar to other application areas cybersecurity professionals are reluctant to black-box ML-based cybersecurity solutions. Keeping themselves one step ahead of the attacker, it is essential for the security analyst to be aware of the internal automatic decision mechanism of the diploid intelligent model and to precisely reason the input data about the model’s outputs. The application of X-AI in cybersecurity could also be a double-edged sword, that is, besides improving security practices, it could also make the intelligent explainable model vulnerable to adversary attacks. This survey provides a comprehensive examination of various XAI-based IDS approaches, evaluating their impact on cybersecurity practices.  \par

The study reveals that different stakeholders in ML-based IDS acquire varying levels and types of interpretability for decision models, with a predominant focus on feature attribution and saliency mapping to comprehend their impact on model decisions. However, a gap exists in understanding the causality and sensitivity effects of attributes in model interpretability. While explanations often showcase the importance of different features, provide meaningful example inputs, visualize decision boundaries, or employ other techniques, there is a need for nuanced and context-aware attribute analysis that demonstrates their connections to real-world domains.  \par

Despite the current significant strides in addressing the need for interpretability, IDSs still have several limitations in terms of complexity, scalability, the trade-off between explainability and accuracy, susceptibility to adversarial attacks, and the lack of standardization. Overcoming these limitations will be crucial for the development of more robust and reliable Explainable IDS solutions. Particularly, the growing concern of adversarial XAI attacks poses a significant challenge, highlighting the need for enhanced security measures to ensure the resilience of interpretability tools against potential threats. Future research efforts should be directed towards fortifying XAI systems, considering the evolving landscape of cybersecurity and the persistent endeavors of adversaries to exploit vulnerabilities.

\section{Acknowledgment}
\label{sec:sec5}
This work was supported by National Priorities Research Program (NPRP) under Grant NPRP13S-0212-200345 from the Qatar National Research Fund
 (a member of Qatar Foundation). The findings herein reflect the work and are solely the responsibility of the authors.

\bibliography{sn-bibliography}

\section*{Author Information}

\begin{minipage}{0.3\textwidth}
    \centering
    \includegraphics[width=0.7\linewidth]{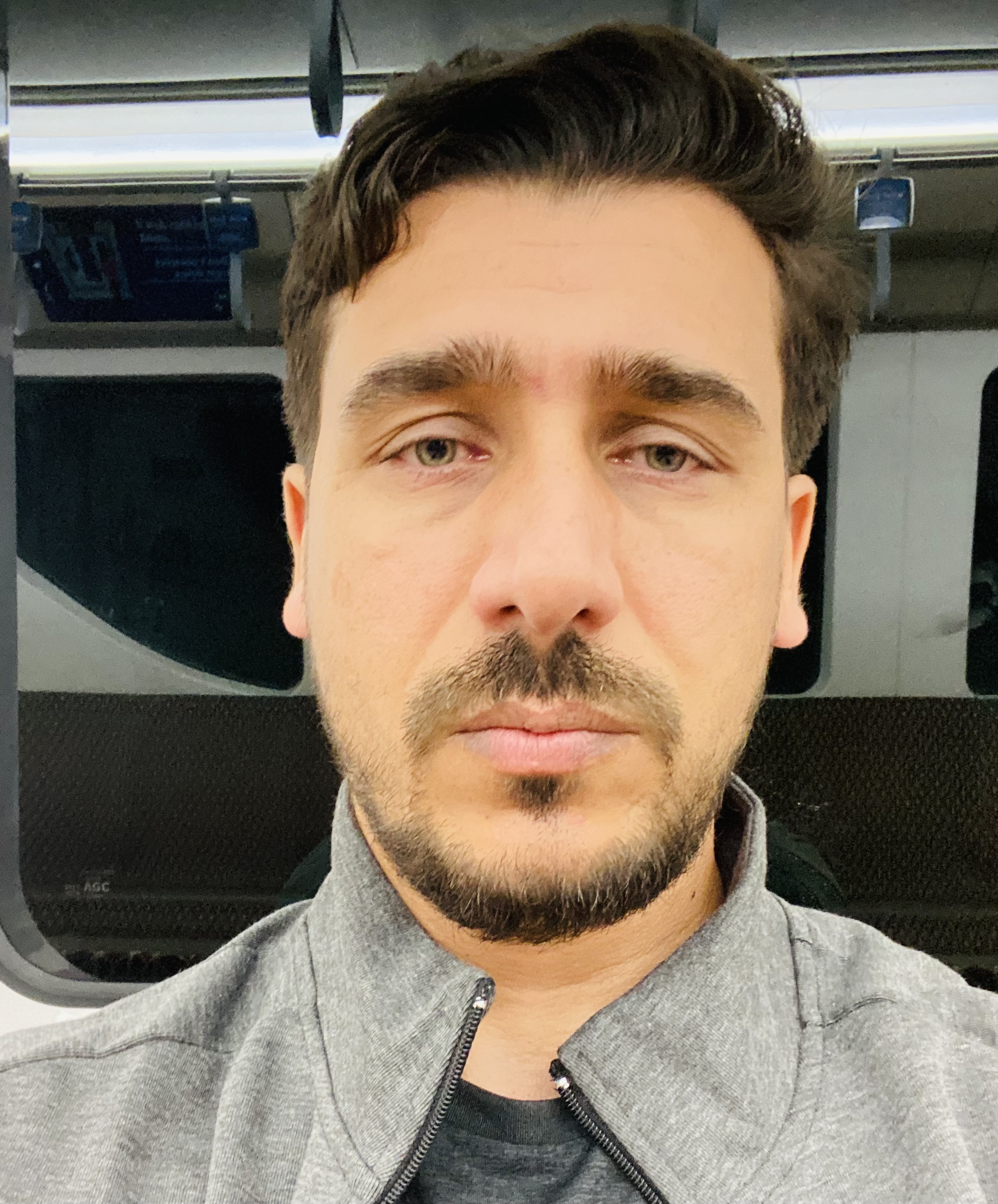}
    \captionsetup{type=figure} 
    \caption*{Naseem Khan}
\end{minipage}%
\hfill
\begin{minipage}{0.8\textwidth}
    \textbf{Naseem Khan} received a bachelor's degree in telecommunication from Hazara University, Mansehra, Pakistan, and master's degrees in Cyber Security from COMSATS University, Abbottabad, Pakistan, in 2016 and 2021, respectively. Currently, he is a Ph.D. student at Hamad bin Khalifa University, Doha, Qatar. His research interests include Cyber Security, specifically malware analysis, intrusion detection and prevention systems, and Deepfake detection. He is working with the Cyber Security Group at the Qatar Computing Research Institute (QCRI).
\end{minipage}

\vspace{1cm} 

\begin{minipage}{0.3\textwidth}
    \centering
    \includegraphics[width=0.8\linewidth]{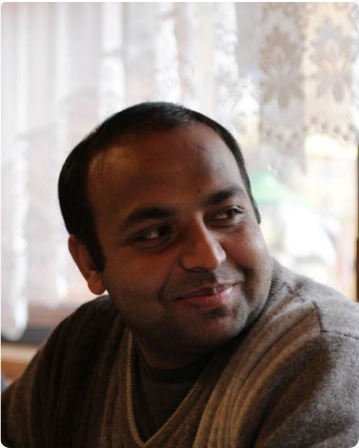}
    \captionsetup{type=figure}
    \caption*{Kashif Ahmad}
\end{minipage}%
\hfill
\begin{minipage}{0.8\textwidth}
    \textbf{Dr. Kashif Ahmad}
     received the bachelor's and master's degrees from the University of Engineering and Technology, Peshawar, Pakistan, in 2010 and 2013, respectively, and the Ph.D. degree from the University of Trento, Italy, in 2017. He worked with the Multimedia Laboratory in DISI, University of Trento. He is currently working as a lecturer in the Department of Computer Science, at Munster Technological University, Cork, Ireland. He has also worked as a Postdoctoral Researcher at Hamad Bin Khalifa University Doha, Qatar, and at ADAPT Centre, Trinity College, Dublin, Ireland. He has authored and co-authored more than 90 journal and conference publications. His research interests include multimedia analysis, computer vision, ML, and signal-processing applications in smart cities. He is a program committee member of several international conferences.\\

\end{minipage}
\vspace{1cm} 

\begin{minipage}{0.3\textwidth}
    \centering
    \includegraphics[width=0.85\linewidth]{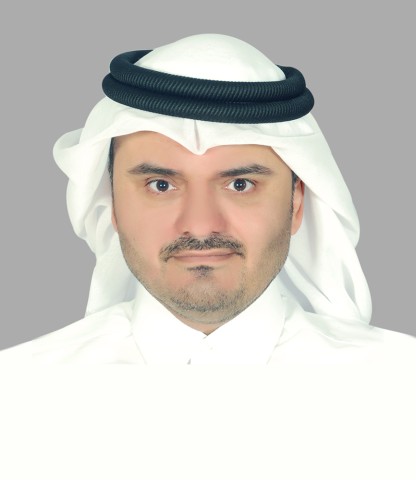}
    \captionsetup{type=figure}
    \caption*{Aref Al-Tamimi}
\end{minipage}%
\hfill
\begin{minipage}{0.8\textwidth}
    \textbf{Dr. Aref Al-Tamimi}
     is a Sr. manager at the national cyber security lab, and Sr. Software Engineer, at Qatar Computing Research Institute (QCRI).
His work includes research and developments of prototypes in Applied research, specifically in anomaly and malicious activity detection in networks.  Before Joining QCRI, He was an entrepreneur in Network and IT solutions in Montreal, Canada. Mr. Al-Tamimi holds a Ph.D. in the field of Internet of Things (IoT) hardware/software security from Hamad Bin Khalifa University (HBKU). He also holds a Masters of Engineering M.E. degree from École Supérieure d'Ingénieurs en Électrotechnique et Électronique (ESIEE Paris) in 2014. He received his Bachelor Degree in Electrical Engineering from Concordia University in Montreal, Canada in 1999. Recently, He has completed the prestigious executive leadership program at Qatar leadership Center(QLC).\\

\end{minipage}
\vspace{1cm} 

\begin{minipage}{0.3\textwidth}
    \centering
    \includegraphics[width=0.8\linewidth]{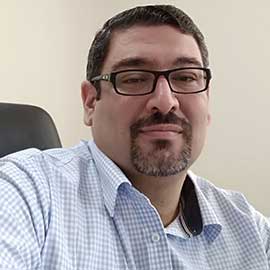}
    \captionsetup{type=figure}
    \caption*{Mohammed M. Alani}
\end{minipage}%
\hfill
\begin{minipage}{0.8\textwidth}
    \textbf{Prof. Mohammed M. Alani}
    holds a PhD in Computer Engineering with specialization in network security. He has worked as a professor, and a cybersecurity expert in many countries around the world. His experience includes working in many academic institutions, network and security consultancies in the Middle-East, and Cybersecurity Program Manager in Toronto Canada. He currently works as a Cybersecurity Professor at Seneca Polytechnic, and a Research Fellow at Toronto Metropolitan University, Toronto, Canada. He is currently serving as an ACM Distinguished Speaker.\\

\end{minipage}
\vspace{1cm} 

\begin{minipage}{0.3\textwidth}
    \centering
    \includegraphics[width=0.8\linewidth]{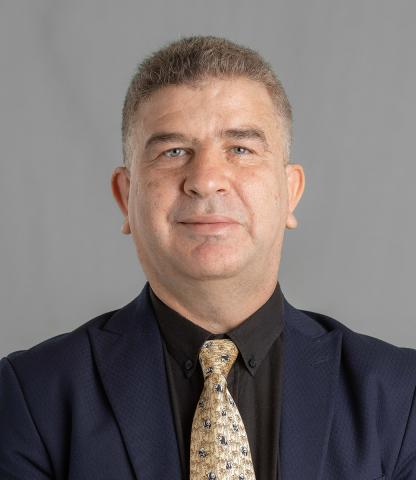}
    \captionsetup{type=figure}
    \caption*{Amine Bermak}
\end{minipage}%
\hfill
\begin{minipage}{0.8\textwidth}
    \textbf{Prof. Amine Bermak}
     (Fellow IEEE), holds a master's and Ph.D. in Electrical and Electronic Engineering from Paul Sabatier University, France. He has held various positions in academia and industry in France, the U.K., Australia, Hong Kong, and is currently a Professor and the Associate Dean at Hamad Bin Khalifa University, Qatar. Prof. Bermak has published over 500 articles, designed 50+ chips, and supervised numerous Ph.D. and M.Phil. students. He has received various prestigious awards for his teaching excellence, including the University Michael G. Gale Medal and the Engineering Teaching Excellence Awards at HKUST. He is also an IEEE Distinguished Lecturer.\\

\end{minipage}

\begin{minipage}{0.3\textwidth}
    \centering
    \includegraphics[width=0.85\linewidth]{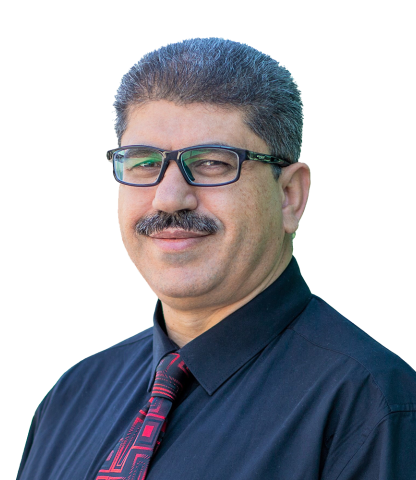}
    \captionsetup{type=figure}
    \caption*{Issa Khalil}
\end{minipage}%
\hfill
\begin{minipage}{0.8\textwidth}
    \textbf{Dr. Issa Khalil}
      obtained a Ph.D. in Computer Engineering from Purdue University in 2007. He is currently a Principal Scientist in the Cyber Security Group at the Qatar Computing Research Institute (QCRI). Prior to joining QCRI, he served as an Associate Professor and Department Head of the Information Security Department at the College of Information Technology (CIT) at the United Arab Emirates University (UAEU). He has more than 27 years of experience in cyber security research, networking, and management.  Dr Khalil has expertise in diverse cyber security areas, including security data analytics, network security, AI/ML security and privacy, cyber threat intelligence, and private data sharing. Dr. Khalil’s research not only results in profound breakthroughs in multiple fronts of cyber security but also helps improve the cyber security practice of the nation through system development, technology transfer and commercialization He is the founding leader of the National Cyber Security Research Lab (NCSRL) and led the design and deployment of one of the largest cyber ranges (CR) in the region.  He participated in developing the Cyber Security Grand Challenge, and the previous and current National Cyber Strategies in Qatar. He has more than 120 top-tier cyber security conferences and journals, and more than 10 US patents and patent disclosures, four of which have been licensed to an international security provider. His achievements in cyber security research and innovations have been featured in many news outlets including BBC and Qatar TV and have been recognized with prestigious awards such as the "Tech Team Award" from Hamad Bin Khalifa University, the HBKU Research Excellence Award, and the Outstanding Professor award for outstanding performance in research, teaching, and service from UAEU.\\

\end{minipage}
\vspace{1cm} 

\end{document}